\documentclass[10pt,journal,twocolumn]{IEEEtran}
\newif\ifCLASSOPTIONromanappendices \CLASSOPTIONromanappendicestrue
\usepackage{tikz}
\usepackage[siunitx]{circuitikz}
\usepackage[utf8]{inputenc}
\usepackage{fontenc}
\usepackage{nicematrix}
\usepackage{a0size}         
\usepackage{amssymb}
\usepackage{amsmath}
\usepackage{empheq}
\usepackage{upgreek}
\usepackage{changepage}
\usepackage{multicol}
\usepackage[english]{babel}     
\usepackage{epsfig}
\usepackage{bm}
\usepackage{oplotsymbl}
\usepackage{bbm}
\usepackage{enumitem}
\usepackage{amsfonts,color,amsthm,amsmath, lscape,graphics}
\usepackage{subfiles}
\usepackage{comment}
\usepackage{algorithm} 
\usepackage{algpseudocode}

\usepackage[font=footnotesize]{caption}
\usepackage[font=footnotesize]{subcaption}

\usepackage[labelformat=simple]{subcaption}
\usepackage{epstopdf}
\usepackage{algorithm} 
\usepackage{algpseudocode}
\usepackage{cite}
\usepackage{booktabs}

\definecolor{awesome}{rgb}{1.0, 0.13, 0.32}
\usepackage{fix2col}

\addto\captionsenglish{}

\usepackage{graphicx}  
\theoremstyle{plain}

\usepackage{multirow} 
\usepackage{relsize}  
\usepackage{textcomp, mathtools}
\usepackage{stfloats}
\usepackage{bm}   
\usepackage{pifont}
\usepackage{etoolbox}
\newtheorem{result}{Result}

\newcommand{\norm}[1]{\left\lVert#1\right\rVert_\textrm{F}}

\makeatletter
\patchcmd{\@maketitle}
  {\addvspace{0.5\baselineskip}\egroup}
  {\addvspace{-1\baselineskip}\egroup}
  {}
  {}
\makeatother
\begin{document}
\title{Super-Wideband Massive MIMO}
\vspace{-1cm}
\author{Mohamed Akrout, Volodymyr Shyianov, Faouzi~Bellili, \IEEEmembership{Member, IEEE},\\ Amine Mezghani, \IEEEmembership{Member, IEEE}, Robert W. Heath, \IEEEmembership{Fellow, IEEE}
\thanks{{The authors are with the Department of Electrical and Computer Engineering (ECE) at the University of Manitoba, Winnipeg, MB, Canada (emails:\{akroutm, shyianov\}@myumanitoba.ca, \{Faouzi.Bellili, Amine.Mezghani\}@umanitoba.ca). R.~W.~Heath is with the ECE department at the North Carolina State University (email: rwheathjr@ncsu.edu). This work was supported by the Discovery Grants Program of the Natural Sciences and Engineering Research Council of Canada (NSERC), Futurewei Technologies, and the US National Science Foundation (NSF) Grant No. ECCS-1711702 \& CNS-1731658. }}}

\maketitle
\begin{abstract}
We present a unified model for connected antenna arrays with a large number of tightly integrated (i.e., coupled) antennas in a compact space within the context of massive multiple-input multiple-output (MIMO) communication. We refer to this system as tightly-coupled massive MIMO. From an information-theoretic perspective, scaling the design of tightly-coupled massive MIMO systems in terms of the number of antennas, the operational bandwidth, and form factor was not addressed in prior art. We investigate this open research problem using a physically consistent modeling approach for far-field (FF) MIMO communication based on multi-port circuit theory. In doing so, we turn mutual coupling (MC) from a foe to a friend of MIMO systems design, thereby challenging a basic percept in antenna systems engineering that promotes MC mitigation/compensation. We show that tight MC widens the operational bandwidth of antenna arrays thereby unleashing a missing MIMO gain that we coin ``bandwidth gain''. Furthermore, we derive analytically the asymptotically optimum spacing-to-antenna-size ratio by establishing a condition for tight coupling in the limit of large-size antenna arrays with quasi-continuous apertures. We also optimize the antenna array size while maximizing the achievable rate under fixed transmit power and inter-element spacing. Then, we study the impact of MC on the achievable rate of MIMO systems under line-of-sight (LoS) and Rayleigh fading channels. These results reveal new insights into the design of tightly-coupled massive antenna arrays as opposed to the widely-adopted ``disconnected'' designs that disregard MC by putting faith in the half-wavelength spacing rule.
\end{abstract}
\section{Introduction}\label{Section 1}
\subsection{Background and Motivation}

Multiple-input multiple-output (MIMO) antenna arrays are ubiquitously used in modern telecommunication systems to increase the achievable rate of wireless transmissions over rich multi-path propagation channels \cite{foschini1998limits,sarkar2008physics}. Their performance analysis has evolved around the seminal work of Shannon \cite{shannon1948mathematical} which laid their narrow-band performance to scale under the asymptotic regime of the signal-to-noise ratio (SNR) and infinite antenna sizes. Since future wireless systems are expected to be super-wideband (i.e., several octaves of bandwidth) spanning both sub-6GHz and mmWave bands, the post-5G MIMO technology requires a drastic shift in the design of antenna arrays in terms of scalability toward higher data rates, broadband capabilities, and the support of massive connectivity with the ever-growing number of internet-of-things (IoT) devices. The reason for the super-wideband requirement is that future antenna systems are expected to be multi-functional (i.e., used for sensing and communication), multi-band, multi-standard and multi-operator as opposed to the current technology \cite{saha2019multifunctional}.

\noindent From this perspective, further research is required to characterize the physical limitations of wireless systems and ultimately put forward new ideas to scale modern communication systems in terms of the number of antennas, the operational bandwidth supported by their antenna circuits while complying with compact form factor requirements. A good example of this is antenna systems in modern smartphones. With the constantly increasing screen-to-body ratio, the antennas should cover multiple (possibly non-contiguous) frequency bands while being confined into physical volumes which cannot be miniaturized beyond the Chu limit \cite{chu1948physical} to avoid a degradation in the operational bandwidth.

\begin{figure*}[t!]
    \centering
    \includegraphics[scale=0.7]{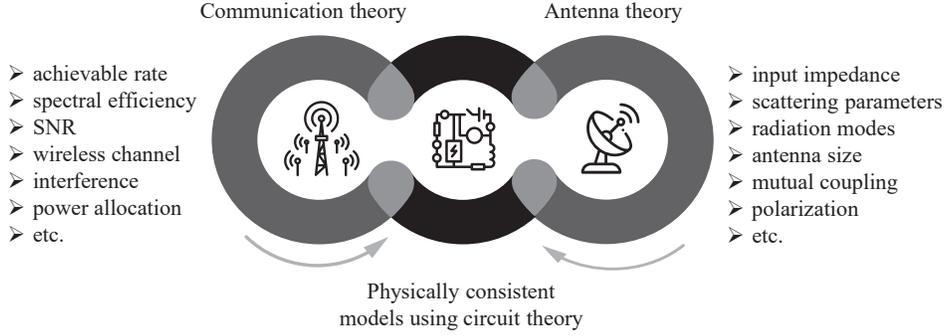}
    \caption{Physically consistent models bridge the gap between communication and antenna theories.}
    \label{fig:comms-antenna}
    \vspace{-0.2cm}
\end{figure*}

To address the scalability of the achievable rate of future MIMO communication systems, it is therefore essential to account for two constraints simultaneously: $i)$ the antenna size constraint due to the size-bandwidth trade-off from Chu's theory \cite{chu1948physical} and $ii)$ the mutual coupling effects (MC) induced by the antenna elements being in close proximity due to the compact form factor requirement \cite{bird2021mutual}. The major difficulty in this problem is that the performance criterion of interest (i.e., the achievable rate) emanates from communication theory while the constraints (i.e., antenna size and mutual coupling) arise from antenna theory. In other words, optimizing the achievable rate of MIMO while taking into account their physical size constraints is a problem that lies at the interface of communication and antenna theories \cite{gabor1953communication}.
Although antennas are fundamental devices for wireless transmissions, the analysis and design of MIMO systems has historically evolved around the basic percept of separating the mathematical abstractions of communication theory\footnote{Particularly the celebrated Shannon capacity formula for band-limited additive white Gaussian noise (AWGN) channels.}\cite{shannon1948mathematical} and the physical design considerations from antenna and electromagnetic theories \cite{balanis,jackson1999classical}. For instance, the wireless community assesses the performance of MIMO systems in terms of achievable rate and spectral efficiency criteria while the figure of merit for antenna design is the scattering parameters. Research effort has been recently made to bridge such assessment gap between communication and antenna communities, e.g., \textit{wave theory of information} \cite{franceschetti2017wave}, \textit{electromagnetic information theory} \cite{gruber2008new, migliore2008electromagnetics}, and \textit{circuit theory for communication} \cite{ivrlavc2010toward}. Most of the existing work, however, is limited to narrow-band communications, and very few \cite{gustafsson2004spectral,shyianov2021achievable} have further imposed the antenna size as the only physical constraint while completely ignoring MC effects. Mutual coupling is actually inevitable between any two closely-spaced antennas being mounted on the same transceiver. Coping with the MC effects is crucial due to the increasing need for packing a massive number of antennas on beyond 5G (B5G) platforms and towers. Indeed, MC can be a foe and a friend at the same time owing to the many advantages it has (e.g., effective current sheet, array calibration) and the various challenges it rises (e.g., beam width/gain decrease, reflection resonances, etc.)  \cite{bird2021mutual}. The study of MC in antenna arrays has heavily focused on decoupling neighboring antennas to mitigate its adverse effects on the experienced SNR \cite{yuan2006performance,wang2016performance}, channel estimation \cite{lu2007effect}, angle of arrival acquisition \cite{lui2010mutual}, to name a few (see \cite{chen2018review} for a comprehensive review). However, little effort has been devoted to investigate its positive impact on the wideband behaviour of antenna arrays \cite{aumann1989phased,wei2015mutual,cavallo2011connected}.

During the recent years, a new approach has emerged for the design of wideband antenna arrays in which MC is intentionally introduced between the antenna elements by reducing their inter-spacing (down to a fraction of the wavelength spacing), i.e., below the widely-adopted half-wavelength spacing in modern wireless MIMO systems \cite{cavallo2011connected}. By doing so, the antenna elements become electrically connected due to MC, thereby yielding a single ``connected array'' periodically fed by a finite number of feeding points. While the antenna elements resonate separately, their tightly coupled behavior provides major advantages such as wideband capabilities and low-level cross-polarization. This is because the MC effects between neighboring elements increase the effective length of the connected antenna array, thereby allowing resonance at larger wavelengths, i.e., lower frequencies \cite{cavallo2011connected}. For this reason, the overall performance of the connected array surpasses the aggregate performance of the isolated resonant antenna elements.

 In this paper, we refer to MIMO systems based on connected arrays as ``tightly-coupled massive MIMO''. Unlike the conventional information-theoretic analysis of MIMO communication systems with decoupled (i.e., half-wavelength spaced) antennas,  we put forward the pivotal role of MC in tightly-coupled massive MIMO as a key enabler for the wideband behaviour that has been recently observed in practical implementations of connected array designs \cite{cavallo2018connected,cavallo20123, van2020wideband}. To this end, we resort to the physically consistent modeling approach of wireless communication systems which bridges the gap between communication and antenna theories as depicted in Fig.~\ref{fig:comms-antenna}. Physically consistent models \cite{ivrlavc2010toward} are circuit-based models describing the wireless communication components using circuit theory analysis. Such models can connect the governing physics of antennas with the mathematics of communication systems using circuit theory without ignoring the wave propagation properties of the channel (e.g., fading, shadowing, scattering). Hence, they provide a powerful modeling tool of the antenna device within the radio-frequency (RF) chain (e.g., matching network, low-noise amplifier). They also incorporate the background noise sources of transmit/receive antennas and offer a precise description of the noise correlation at the receiver stemming from the MC effects. For this reason, physically consistent models are an effective abstraction to analyze the realistic physical limits on the achievable rate and to optimize the wideband/compact antenna design \cite{saab2019capacity,saab2019capacitybased,akrout2021achievable}.
\subsection{Contributions}

In this paper, we present \textit{tightly-coupled massive MIMO} as a future promising solution that scales $i)$ in the number and size of antennas by making use of electrically connected small antennas, and $ii)$ in the operational bandwidth which can be widened by packing more antenna elements while harnessing their mutual coupling. From this perspective, tightly-coupled massive MIMO enables the convergence and integration of several radio infrastructures (e.g., WIFI, cellular, satellite, radar) using a common antenna array that is capable to operate/resonate over a wide range of frequency bands. It also presents a new framework for analyzing and creating future sustainable unified radio infrastructure with lower cost/power and environmental footprint.

Toward this goal, we start by combining Chu's theory \cite{chu1948physical} with canonical minimum scattering (CMS) antennas \cite{kahn1965minimum} to develop a physically-consistent description of the wireless tightly-coupled massive MIMO communication channel. This allows one to jointly account for antenna size limitations using Chu's theory and MC effects \textit{independently of the antenna type}. By operating at the lowest transverse magnetic (TM) radiation mode only, Chu's CMS antennas have the widest operational bandwidth compared to all other types of Chu’s antennas (that are excited by higher-order radiation modes). Hence, our analysis captures the salient features of the broadest possible antenna that can be realized within a prescribed sphere, i.e., with no specific antenna design in mind. Using the mutual coupling model between two Chu's CMS antennas \cite{akrout2021achievable}, we obtain the impedances of transmit and receive arrays of Chu's CMS antenna elements, each of which being confined in a spherical volume of radius $a_{\text{T}}$ and $a_{\text{R}}$, respectively. We also derive the far-field (FF) transimpedance matrices for both line-of-sight and Rayleigh fading MIMO channels.

\begin{table*}[bp]
    \centering
    \caption{{\fontsize{11pt}{11pt}Summary of abbreviations.}}
    \small
    \begin{tabular}{cc||cc}
        \hline
        FF & Far-Field & CMS & Canonical Minimum Scattering \\ \hline
        NF & Near-Field & AWGN &  Additive White Gaussian Noise  \\ \hline
        MC & Mutual Coupling & LoS & Line-of-sight  \\ \hline
        MIMO & Multiple-Input-Multiple-Output & PA & Power Allocation  \\ \hline
        SIMO & Single-Input-Multiple-Output & SNR & signal-to-noise ratio  \\ \hline
        MISO & Multiple-Input-Single-Output & LNA & Low-Noise Amplifier  \\ \hline
        SISO & Single-Input-Single-Output & ULA & Uniform Linear Array  \\
         \hline
    \end{tabular}
    \label{table:abbreviations}
\end{table*}

Based on the established equivalent circuit models, we then derive the mutual information between the input and output signals of the MIMO system under MC and the antenna size constraint. To the best of our knowledge, our work is the first of its kind to analyze tightly-coupled massive MIMO systems under realistic physical limitations from antenna and applied electromagnetic theories. We recognize the well-known concepts of multi-port network modelling \cite{ivrlavc2010toward} and achievable rate optimization \cite{wallace2004mutual} as the adequate framework to integrate/optimize new dimensions of analysis (i.e., antenna spacing, antenna size constraint from Chu's theory and antenna orientation from Chu-Hertz equivalence) into the analysis of multi-port circuit for communication. Further, our investigation reveals for the very first time the limitation of the half-wavelength antenna spacing that has so far become an orthodox rule in antenna array design. Specifically, it highlights the following two main results:
\begin{itemize}[leftmargin=*]
    \item the MC effect within the transmit/receive colinearly connected antenna arrays is key to widening the operational bandwidth, thereby leading to a bandwidth gain on top of the conventional multiplexing/antenna/diversity gains of MIMO systems. The bandwidth gain enables the promised broadband capabilities of future MIMO communication\footnote{This is fundamentally different from the practically inaccessible super gain of antenna arrays where higher array gain is obtained at the cost of a very narrow bandwidth.}. It also implies that the number of degrees of freedom of a broadband communication system increases with its bandwidth gain. This adds to the conventional antenna/multiplexing gains and is to be opposed to narrow-band systems where only spatial degrees of freedom have been considered \cite{tse2005fundamentals}. While this result is known within the radar community, the achievable rate formulation provides us with a precise characterization of the trade-off in terms of antenna size, frequency-dependent power allocation, achievable rate, and communication range.
    
    \item We determine analytically the asymptotically optimum spacing-to-antenna-size ratio for tight coupling for Chu's CMS antenna arrays with quasi-continuous aperture. We also numerically show the existence of an optimum spacing-to-antenna-size ratio (corresponding to an optimum mutual coupling) for finite arrays that converges to the asymptotic one as the number of antenna elements increases. The corresponding optimum inter-element spacing suggests that the antenna elements shall be connected and tightly coupled for broadband behaviour.
\end{itemize}

\subsection{Organization of the Paper, Notations, and Abbreviations}
We structure the rest of this paper as follows. In Section~\ref{sec:preliminaries}, we introduce the relevant preliminaries of the physically consistent modeling at the interface of communication and antenna theories by reviewing the following concepts: circuit theory for MIMO communication, Chu's theory \cite{chu1948physical} for size-constrained antenna characterization and the mutual coupling model between two Chu's CMS antennas \cite{akrout2021achievable}. In Section~\ref{sec:mimo-system}, we present the circuit-theoretic MIMO communication model and derive its input-output relationship for an arbitrary joint impedance matrix $\bm{Z}_{\text{MIMO}}$. We also consider the FF transimpedance between any pair of transmit and receive antennas when the latter are modeled as Chu's antennas, thereby accounting for the physical antenna size constraint. In Section \ref{sec:mutual-self-impedance-computation}, we derive both the transimpedance matrix for FF communications and the transmit/receive impedance matrices for colinear and parallel uniform linear antenna arrays (ULAs). Finally, our simulation results are presented in Section~\ref{sec:results} for both colinear and parallel antenna orientations, from which we draw out some concluding remarks regarding the importance of MC in enabling broadband capabilities of future wireless MIMO systems.
\newline 
\indent The following notation is adopted throughout the paper. Given any complex number,  $\Re\{\cdot\}$, $\Im\{\cdot\}$, and $\{\cdot\}^*$  return its real part, imaginary part, and complex conjugate, respectively. Given any matrix $\bm{A}$, $\bm{A}^{\textsf{T}}$ and $\bm{A}^{\textsf{H}}$ refer to its transpose and hermitian, and $\text{diag}(\bm{A})$ returns a diagonal matrix by setting the off-diagonal elements of $\bm{A}$ to zero. The statistical expectation and variance are denoted as $\mathbb{E}[\cdot]$ and $\textrm{Var}[\cdot]$, respectively. We also use the non-italic letter $\textrm{j}$ to denote the imaginary unit (i.e., $\textrm{j}^{2}=-1$). Throughout the paper, $c$ denotes the speed of light in vacuum (i.e., $c \approx 3\times10^8$), $\lambda$ is the wavelength, and $k_b = 1.38 \times 10^{-23}\, \mathrm{m}^{2}\, \mathrm{kg} \,\mathrm{s}^{-2}\, \mathrm{K}^{-1}$ is the Boltzmann constant. Finally, $k_0$ and $\eta_0$ are the wave number and the wave impedance in free space.

\section{Preliminaries}\label{sec:preliminaries}
\subsection{Circuit theory for MIMO communication}
The analysis of a real pass-band signal, $\mathbf{\textsf{v}}(t)$, is more convenient to be carried out in the frequency-domain by defining its Fourier transform, $\bm{v}(f)$, as follows
\begin{equation}
    \bm{v}(f) = \int_{-\infty}^{+\infty}{\mathbf{\textsf{v}}(t)\,e^{-\textrm{j}2{\pi}ft}}\,\text{d}t.
\end{equation}
When $\mathbf{\textsf{v}}(t)$ is random, however, it is more convenient to use its Fourier Transform truncated to some interval of time $T_0$:
\begin{equation}\label{trancated_Fourier}
    \bm{v}_{T_0}(f) = \int_{-\frac{T_0}{2}}^{+\frac{T_0}{2}}{\mathbf{\textsf{v}}(t)\,e^{-\textrm{j}2{\pi}ft}}\,\text{d}t.
\end{equation}
The study of random signals transmitted through a wireless channel requires a circuit-theoretic analysis that is consistent with the governing laws of physics \cite{ivrlavc2010toward}. In this respect, transmitted/received signals are either voltages or currents flowing through the feeding ports of the transmit/receive antennas. Therefore, characterizing the MIMO relationship between port variables at the transmitter(s) and receiver(s) is all that is needed to consistently model both near-field and far-field MIMO communication channels.\vspace{-0.3cm}

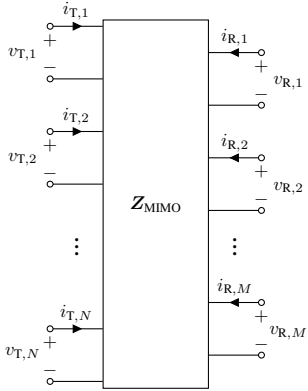
\begin{figure}[h!]
    \centering
    \hspace{-2cm}
    \begin{adjustwidth*}{}{2.5cm} 
\begin{circuitikz}[american voltages, american currents, scale=0.7, transform shape]
\draw (0,0)
node[draw,minimum width=2cm,minimum height=7cm] (load) {$\bm{Z}_{\textrm{MIMO}}$};
\draw ($(load.west)!0.75!(load.north west)+(0,0.75)$) coordinate (I1);
\draw ($(load.west)!0.75!(load.north west)+(0,-0.25)$) coordinate (I1back);
\draw ($(load.west)!0.75!(load.north west)+(0,-1.25)$) coordinate (I2);
\draw ($(load.west)!0.75!(load.north west)+(0,-2.25)$) coordinate (I2back);
\draw ($(load.west)!0.75!(load.north west)+(0,-5)$) coordinate (IM);
\draw ($(load.west)!0.75!(load.north west)+(0,-6)$) coordinate (IMback);
\draw ($(load.east)!0.75!(load.north east)+(0,0.75)$) coordinate (Ir1);
\draw ($(load.east)!0.75!(load.north east)+(0,-0.25)$) coordinate (Ir1back);
\draw ($(load.east)!0.75!(load.north east)+(0,-1.25)$) coordinate (Ir2);
\draw ($(load.east)!0.75!(load.north east)+(0,-2.25)$) coordinate (Ir2back);
\draw ($(load.east)!0.75!(load.north east)+(0,-5)$) coordinate (IrM);
\draw ($(load.east)!0.75!(load.north east)+(0,-6)$) coordinate (IrMback);
\draw ($(I1) + (-1,0)$) to[short,i>^=$i_{\text{T},1}$,o-] (I1);
\draw ($(I1back) + (-1,0)$) to[short,o-] (I1back);
\draw ($(I2) + (-1,0)$) to[short,i>^=$i_{\text{T},2}$,o-] (I2);
\draw ($(I2back) + (-1,0)$) to[short,o-] (I2back);
\draw ($(IM) + (-1,0)$) to[short,i>^=$i_{\text{T},N}$,o-] (IM);
\draw ($(IMback) + (-1,0)$) to[short,o-] (IMback);
\draw ($(Ir1) + (1,-0.5)$) to[short,i>_=$i_{\text{R},1}$,o-] ($(Ir1)+ (0,-0.5)$);
\draw ($(Ir1back) + (1,-0.5)$) to[short,o-] ($(Ir1back)+ (0,-0.5)$);
\draw ($(Ir2) + (1,-0.5)$) to[short,i>_=$i_{\text{R},2}$,o-] ($(Ir2) + (0, -0.5)$);
\draw ($(Ir2back) + (1,-0.5)$) to[short,o-] ($(Ir2back) + (0, -0.5)$);
\draw ($(IrM) + (1,0.5)$) to[short,i>_=$i_{\text{R},M}$,o-] ($(IrM) + (0,0.5)$);
\draw ($(IrMback) + (1,0.5)$) to[short,o-] ($(IrMback) + (0,0.5)$);
\draw ($(I1) + (-1,-0.15)$) to [open,v={\small }] ($(I1) + (-1,-0.9)$);
\draw ($(I2) + (-1,-0.15)$) to [open,v={\small }] ($(I2) + (-1,-0.9)$);
\draw ($(IM) + (-1,-0.15)$) to [open,v={\small }] ($(IM) + (-1,-0.9)$);
\draw ($(Ir1) + (1,-0.15-0.5)$) to [open,v={\small }] ($(Ir1) + (1,-0.9-0.5)$);
\draw ($(Ir2) + (1,-0.15-0.5)$) to [open,v={\small }] ($(Ir2) + (1,-0.9-0.5)$);
\draw ($(IrM) + (1,-0.15+0.5)$) to [open,v={\small }] ($(IrM) + (1,-0.9+0.5)$);
\node[] at (-1.5,-0.7) {{\textbf{$\vdots$}}};
\node[] at (1.5,-0.7) {{\textbf{$\vdots$}}};
\node[] at (-2.4,2.85) {{$v_{\text{T},1}~~$}};
\node[] at (-2.4,0.85) {{$v_{\text{T},2}~~$}};
\node[] at (-2.4,-2.85) {{$v_{\text{T},N}~~$}};
\node[] at (2.4,2.85-0.55) {{$~~v_{\text{R},1}$}};
\node[] at (2.4,0.85-0.55) {{$~~v_{\text{R},2}$}};
\node[] at (2.4,-2.85+0.4) {{$~~v_{\text{R},M}$}};
\end{circuitikz}
\end{adjustwidth*}
    \caption{Equivalent circuit-theoretic model for MIMO communication channels.}
    \label{fig:mimo-channel-circuit-models}
\end{figure}

\noindent As depicted in Fig. \ref{fig:mimo-channel-circuit-models}, a MIMO communication channel is modeled as a \textit{multi-port network} with ($N+M$) ports representing $N$ transmit and $M$ receive antennas. The voltage (resp. current) vectors $\bm{v}_{\textrm{T}}(f) = \big[ v_{\text{T},1}(f), v_{\text{T},2}(f), \dots, v_{\text{T},N}(f)\big]^\top$ \big(resp. $\bm{i}_{\textrm{T}}(f) = \big[ i_{\text{T},1}(f), i_{\text{T},2}(f), \dots, i_{\text{T},N}(f)\big]^\top$\big) and $\bm{v}_{\textrm{R}}(f) = \big[ v_{\text{R},1}(f), v_{\text{R},2}(f), \dots, v_{\text{R},M}(f)\big]^\top$ \big(resp. $\bm{i}_{\textrm{R}}(f) =  i_{\text{R},1}(f), i_{\text{R},2}(f), \dots, i_{\text{R},M}(f)\big]^\top$\big) denote the voltages (resp. currents) at all the $N$ transmit and $M$ receive antennas, respectively. The relationship between the transmit/receive voltages and currents is given by Ohm's law:
\begin{equation}\label{eq:mimo-V=ZI}
    \Bigg[\begin{array}{l}
\bm{v}_{\textrm{T}}(f) \\
\bm{v}_{\textrm{R}}(f)
\end{array}\Bigg]~=~\underbrace{\Bigg[\begin{array}{cc}
\bm{Z}_{\text{T}}(f) & \bm{Z}_{\text{TR}}(f)\\
\bm{Z}_{\text{RT}}(f) & \bm{Z}_{\text{R}}(f)
\end{array}\Bigg]}_{\bm{Z}_{\text{MIMO}}}\,\Bigg[\begin{array}{l}
\bm{i}_{\textrm{T}}(f) \\
\bm{i}_{\textrm{R}}(f)
\end{array}\Bigg].
\end{equation}

\noindent The $N\times N$ and $M\times M$ impedance matrices $\bm{Z}_{\text{T}}$ and  $\bm{Z}_{\text{R}}$ stand for the impedances of the transmit and receive arrays, respectively. Their diagonal elements correspond to the  transmit/receive self-impedances (i.e., when all other antennas are absent) and their off-diagonal elements represent the mutual impedances between the antenna elements within the transmit/receive array. On the other hand, the matrices $\bm{Z}_{\text{RT}}$ and  $\bm{Z}_{\text{TR}}$ are $M \times N$ and $N \times M$ symmetric matrices that gathers the pairwise transimpedances between the transmit-receive and receive-transmit antenna elements, respectively. Since antennas are reciprocal devices, it follows that $\bm{Z}_{\text{\textrm{RT}}} = \bm{Z}_{\text{\textrm{TR}}}^\top$ \cite{balanis}. Together, the four aforementioned matrix blocks, $\bm{Z}_{\text{\textrm{T}}}$, $\bm{Z}_{\text{\textrm{R}}}$, $\bm{Z}_{\text{\textrm{RT}}}$, and $\bm{Z}_{\text{\textrm{TR}}}$ form the symmetric $(N+M) \times (N+M)$ joint impedance matrix $\bm{Z}_{\text{MIMO}}$ of the MIMO system. The knowledge of $\bm{Z}_{\text{MIMO}}$ fully characterizes the MIMO communication system and enable its analysis with a much more convenient and compact multi-port matrix description.

\noindent For FF communications, the signal attenuation in the FF region between the transmitter and the receiver is very large, i.e.:
\begin{equation}\label{eq:unilateral-approx}
     \norm{\bm{Z}_{\textrm{TR}}(f)} ~=~ \norm{\bm{Z}_{\textrm{RT}}(f)} ~\ll~ \min\big(\norm{\bm{Z}_{\textrm{T}}(f)}, \norm{\bm{Z}_{\textrm{R}}(f)}\big).
\end{equation}
This justifies the so-called ``unilateral approximation'' \cite{ivrlavc2010toward} which stipulates that the receive-transmit transimpedance $\bm{Z}_{\textrm{TR}}(f)$ can be ignored (i.e., $ \bm{Z}_{\textrm{TR}}(f) \approx \mathbf{0}$) in the FF region wherein only the transmit antennas influence the electromagnetic properties of the receive antennas.

\begin{figure*}[t]
\begin{equation}\label{eq:mutual-coupling-SISO}
    \begin{aligned}[b]
        &Z_{\text{TR}}(f) = Z_{\text{RT}}(f)  =-3\,\sqrt{\Re\big[Z_{\text{T}}(f)\big]\,\Re\big[Z_\text{R}(f)\big]} \Bigg[\frac{1}{2}\,\sin(\beta)\,\sin(\gamma)\bigg(\frac{1}{\textrm{j}k_0d} + \frac{1}{(\textrm{j}k_0d)^2} +\frac{1}{(\textrm{j}k_0d)^3}\bigg)\\
        &\hspace{3.5cm}+\cos(\gamma)\,\cos(\beta)\bigg(\frac{1}{(\textrm{j}k_0d)^2} + \frac{1}{(\textrm{j}k_0d)^3}\bigg)\Bigg]\,e^{-\textrm{j}k_0d} ~\triangleq~ \mathcal{MC}(d,\beta,\gamma),
    \end{aligned}
\end{equation}
\vspace{-0.1cm}
\rule{\textwidth}{0.4 pt}
\vspace{-0.4cm}
\end{figure*}

\subsection{The antenna size constraint}\label{subsec:antenna-chu-constraint}
Since the maximum gain of an antenna is proportional to its physical size compared to wavelength \cite{balanis}, it is tempting to argue that the achievable rate of a MIMO system should also be proportional to its transmit/receive antenna array sizes. To confirm or refute this statement, one should resort to Chu's theory \cite{chu1948physical} so as to perform the analysis using the most efficient antenna structure that can be designed under a physical size constraint. Indeed, the seminal work of Chu \cite{chu1948physical} laid the foundations for the equivalent circuit models of any antenna whose structure can be embedded inside a spherical volume of a given radius $a$. More specifically, Chu derived an equivalent circuit network for each $n^{\text{th}}$ spherical $\text{TM}_{n}$ radiation mode of a given electromagnetic field in free space. For broadband communication applications, it suffices to consider antennas having the first radiation mode only (i.e., $n = 1$) since they have $i)$ the lowest Q-factor or equivalently the broadest bandwidth\footnote{A larger bandwidth could be achieved (i.e., improved Chu limit) if we combine TM$_1$ and TE$_1$, known as magneto-electric antenna \cite{hansen2011small}. For simplicity consideration, in this work we consider electric antennas only.}, and $ii)$ a gain of $3/2$ in the equatorial plane \cite[chapter 6]{harrington1961pp}. These are CMS antennas and hence have closed-form mutual impedances in MIMO settings. Throughout the rest of this paper, we will call these antennas as ``Chu antennas''. 
For this reason, we consider the equivalent voltage $V_{1}$ and current $I_{1}$ of a Chu antenna which are given by
\cite{chu1948physical}:
\begin{subequations}\label{appendix-eq:V-I-TM1}
    \begin{align}
    V_{1}&~=~\sqrt{\frac{8 \pi \eta_0}{3}} \,\frac{A_{1}}{k_0} \, \bigg(1 + \frac{1}{\textrm{j}k_0a} - \frac{1}{\textrm{j}(k_0a)^2}\bigg)\,e^{-\textrm{j}k_0a}~[\text{V}], \\
I_{1}&~=~-\sqrt{\frac{8 \pi\eta_0}{3}}\,\frac{A_{1}}{k_0} \, \bigg(1 + \frac{1}{\textrm{j}k_0a}\bigg) \,e^{-\textrm{j}k_0a}~[\text{A}].
    \end{align}
\end{subequations}
where $A_1$ is the $\textrm{TM}_1$ mode's complex coefficient. The equivalent ladder circuit network for the TM$_1$ mode is illustrated in Fig.~\ref{fig:tm1}. Using basic circuit analysis, one can write the self/input impedance $Z_\text{Chu}$ as:
\begin{figure}[h!]
\centering
\begin{circuitikz}[american voltages, american currents, scale=0.7, every node/.style={transform shape}]
\draw (0,0) node[anchor=east]{}
 to[short, o-*] (3,0);
 \draw (3,2) to[L, label=\mbox{\small{$L=\frac{a \,R}{c}$}}, *-*] (3,0);
 \draw (3,0) -- (5,0);
 \draw (5,2) to[/tikz/circuitikz/bipoles/length=30pt, R, l=\mbox{\small{$R$}}, -] (5,0);
 \draw (3,2) -- (5,2);
 \draw (0,2) node[anchor=east]{}
  to[C, i>_=\mbox{\small{$I_1(f)$}}, label=\mbox{\small{$C=\frac{a}{cR}$}}, o-*] (3,2);
  \draw[-latex] (1,0.5) -- node[above=0.05mm] {$Z_\text{Chu}(f)$} (2, 0.5);
  \draw (0,2) to [open,v=\small{$V_1(f)$}] (0,0);
\end{circuitikz}
\caption{Equivalent circuit of the TM$_1$ radiation mode of a Chu's antenna.}
\label{fig:tm1}
\vspace{-0.5cm}
\end{figure}
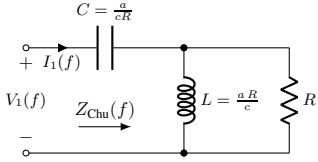
\begin{equation}\label{eq:Z1-chu}
\begin{aligned}
    Z_\text{Chu}(f) ~=~ \frac{V_1(f)}{I_1(f)}~&=~ \underbrace{\frac{1}{\textrm{j}2\pi f\frac{a}{cR}}}_{Z_C(f)} + \underbrace{\frac{1}{\frac{1}{\textrm{j}2\pi f\frac{aR}{c}}+\frac{1}{R}}}_{Z_{\rm L}(f) \,\parallel\, Z_R(f)} \\
    &=~ \frac{c^2R + \textrm{j}2\pi fcaR - (2\pi fa)^2R}{\textrm{j}2\pi f ca - (2 \pi fa)^2}\,\, [\Omega].
\end{aligned}
\end{equation}
 
\noindent As a consequence, Chu antennas are not only the appropriate antenna structures due to their desirable broadband behaviour, but also they are the ones that comply with the specified physical size constraint (i.e., confined in a sphere of a certain radius $a$). From this perspective, this work will develop a MIMO circuit-theoretic model using Chu antennas to obtain the maximum achievable rate for FF MIMO communication with transmit and receive arrays composed of antenna elements of fixed sizes $a_\text{T}$ and $a_\text{R}$, respectively.
 
 \subsection{Mutual coupling model between two Chu's CMS antennas}\label{subsec:mutual-coupling-CMS-Chu}
When the mutual coupling between antenna elements cannot be ignored due to the limited space, the mutual impedance is a practical measure of such proximity effects \cite[pp. 1799–1811]{schelkunoff1952antennas}.
For finite-size MIMO arrays, accounting for the mutual coupling effect has been approached in three different ways \cite{bird2021mutual}:
\begin{itemize}[leftmargin=*]
    \item[$1)$] using forward solvers based on numerical methods such as finite element/difference and the method of moments,
    \item[$2)$] using element-by-element methods based on the construction of an impedance or admittance matrix by means of appropriate antenna terminations,
    \item[$3)$] using minimum scattering methods applied on ``minimum scattering antennas'' \cite{kahn1965minimum} whose electromagnetic properties can be expressed explicitly in terms of their radiation patterns only under appropriate antenna terminations, e.g., open-circuit termination for canonical minimum scattering (CMS) antennas.
\end{itemize}

\begin{figure}[h!]
    \centering
    \includegraphics[scale=0.8]{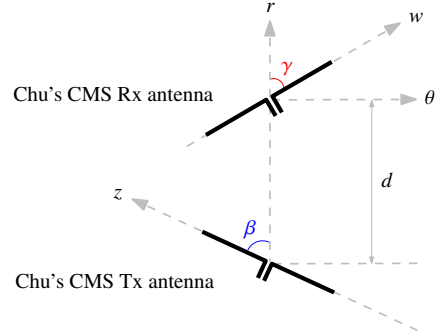}
    \caption{Two Chu's CMS antennas in the same plane, arbitrarily oriented in 2-dimensional free space, and separated by a distance $d$ [m].}
    \label{fig:two-hertz-antennas}
    \vspace{-0.1cm}
\end{figure}

\noindent For CMS antennas, the mutual impedance can be calculated analytically using the ``induced EMF'' method \cite[chapter 25]{orfanidis2002electromagnetic}. Based on a radiated power equivalence between Hertz dipoles and Chu's CMS antennas, it has been recently shown in \cite{akrout2021achievable} that the mutual impedance between two Chu's CMS antennas prescribed within non-overlapping spheres with self-impedances $Z_{\textrm{T}}(f)$ and  $Z_{\text{R}}(f)$ and separated by a distance $d$ is given by (\ref{eq:mutual-coupling-SISO}). There, the angles $\beta$ and $\gamma$ represent the rotation of the dipoles with respect to (w.r.t.) their connecting axis $r$ as depicted in Fig.~\ref{fig:two-hertz-antennas}. Throughout the paper, we refer to the mutual impedance expression in (\ref{eq:mutual-coupling-SISO}), parameterized by the parameter triplet ($d$, $\beta$, $\gamma$), as ``the mutual coupling model $\mathcal{MC}(d, \beta, \gamma)$'' between two Chu's CMS antennas.\vspace{-0.4cm}

\noindent The coupling model in (\ref{eq:mutual-coupling-SISO}) involves two key antenna parameters that are commonly neglected in prior work. These are the antenna size owing to Chu's theory and the relative orientation between the Chu's CMS antennas. Indeed, existing coupling models are based on Bessel functions and assume the antenna radiation to be either isotropic or non-isotropic while disregarding the antenna size constraint \cite{bird2021mutual}. These simplistic models do not account for the mutual coupling effects between non-isotropic radiators (e.g., dipoles) as a function of their size and/or relative orientations.

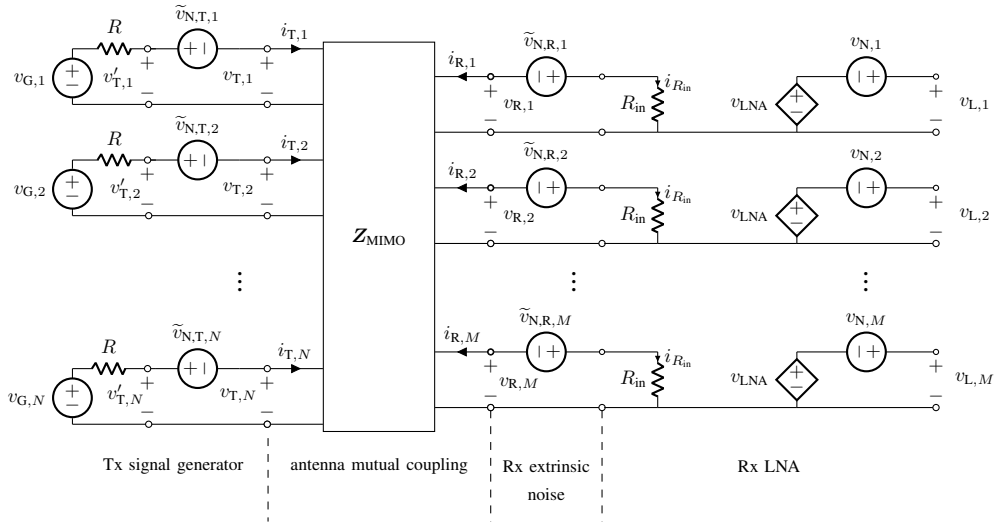
\begin{figure*}[bp]
\vspace{-0.3cm}
\centering
  \begin{circuitikz}[american voltages, american currents, scale=0.74, transform shape]
\draw (0,0)
node[draw,minimum width=2cm,minimum height=7cm] (load) {$\bm{Z}_{\textrm{MIMO}}$}; 
\draw ($(load.west)!0.75!(load.north west)+(0,0.75)$) coordinate (I1);
\draw ($(load.west)!0.75!(load.north west)+(0,-0.25)$) coordinate (I1back);
\draw ($(load.west)!0.75!(load.north west)+(0,-1.25)$) coordinate (I2);
\draw ($(load.west)!0.75!(load.north west)+(0,-2.25)$) coordinate (I2back);
\draw ($(load.west)!0.75!(load.north west)+(0,-5)$) coordinate (IM);
\draw ($(load.west)!0.75!(load.north west)+(0,-6)$) coordinate (IMback);
\draw ($(load.east)!0.75!(load.north east)+(0,0.75)$) coordinate (Ir1);
\draw ($(load.east)!0.75!(load.north east)+(0,-0.25)$) coordinate (Ir1back);
\draw ($(load.east)!0.75!(load.north east)+(0,-1.25)$) coordinate (Ir2);
\draw ($(load.east)!0.75!(load.north east)+(0,-2.25)$) coordinate (Ir2back);
\draw ($(load.east)!0.75!(load.north east)+(0,-5)$) coordinate (IrM);
\draw ($(load.east)!0.75!(load.north east)+(0,-6)$) coordinate (IrMback);
\draw ($(I1) + (-1,0)$) to[short,i>^=$i_{\text{T},1}$,o-] (I1);
\draw ($(I1back) + (-1,0)$) to[short,o-] (I1back);
\draw ($(I2) + (-1,0)$) to[short,i>^=$i_{\text{T},2}$,o-] (I2);
\draw ($(I2back) + (-1,0)$) to[short,o-] (I2back);
\draw ($(IM) + (-1,0)$) to[short,i>^=$i_{\text{T},N}$,o-] (IM);
\draw ($(IMback) + (-1,0)$) to[short,o-] (IMback);
\draw ($(Ir1) + (1,-0.5)$) to[short,i>_=$i_{\text{R},1}$,o-] ($(Ir1back) + (0,0.5)$);
\draw ($(Ir1back) + (1,-0.5)$) to[short,o-] ($(Ir1back) + (0,-0.5)$);
\draw ($(Ir2) + (1,-0.5)$) to[short,i>_=$i_{\text{R},2}$,o-] ($(Ir2back) + (0,0.5)$);
\draw ($(Ir2back) + (1,-0.5)$) to[short,o-] ($(Ir2back) + (0,-0.5)$);
\draw ($(IrM) + (1,0.3)$) to[short,i>_=$i_{\text{R},M}$,o-] ($(IrM) + (0,0.3)$);
\draw ($(IrMback) + (1,0.3)$) to[short,o-] ($(IrMback) + (0,0.3)$);
\draw ($(I1) + (-1,-0.15)$) to [open,v={\small }] ($(I1) + (-1,-0.9)$);
\draw ($(I2) + (-1,-0.15)$) to [open,v={\small }] ($(I2) + (-1,-0.9)$);
\draw ($(IM) + (-1,-0.15)$) to [open,v={\small }] ($(IM) + (-1,-0.9)$);
\draw ($(Ir1) + (1,-0.65)$) to [open,v={\small }] ($(Ir1) + (1,-1.4)$);
\draw ($(Ir2) + (1,-0.65)$) to [open,v={\small }] ($(Ir2) + (1,-1.4)$);
\draw ($(IrM) + (1,0.15)$) to [open,v={\small }] ($(IrM) + (1,-0.6)$);
\draw ($(I1) + (-3.15,-0.15)$) to [open,v={\small }] ($(I1) + (-3.15,-0.9)$);
\draw ($(I2) + (-3.15,-0.15)$) to [open,v={\small }] ($(I2) + (-3.15,-0.9)$);
\draw ($(IM) + (-3.15,-0.15)$) to [open,v={\small }] ($(IM) + (-3.15,-0.9)$);

\draw ($(I1) + (-4.5,-0.15)$) to[/tikz/circuitikz/bipoles/length=33pt, V, l_=$v_{\textrm{G},1}$] ($(I1) + (-4.5,-0.9)$);
\draw ($(I1back) + (-4.5,0)$) to[short,-o] ($(I1back) + (-3.13,0)$);
\draw ($(I1back) + (-3.08,0)$) to[short] ($(I1back) + (-1.05,0)$);
\draw ($(I1back) + (-4.5,0)$) to ($(I1) + (-4.5,-0.9)$);
\draw ($(I1) + (-4.5,-0.15)$) to ($(I1) + (-4.5,0)$);
\draw ($(I1) + (-4.5,0)$)  to[/tikz/circuitikz/bipoles/length=20pt,R,l=$R$,-] ($(I1)+ (-3,0)$);
\draw ($(I1) + (-3,0)$) to[/tikz/circuitikz/bipoles/length=33pt, V, l^=$\widetilde{v}_{\textrm{N,T},1}$, -] ($(I1) + (-1.5,0)$);
\draw ($(I1) + (-3.15,0)$) to[short, o-] ($(I1) + (-3,0)$);
\draw ($(I1) + (-1.5,0)$) to[short] ($(I1) + (-1.05,0)$);
\draw ($(I2) + (-4.5,-0.15)$) to[/tikz/circuitikz/bipoles/length=33pt, V, l_=$v_{\textrm{G},2}$] ($(I2) + (-4.5,-0.9)$);
\draw ($(I2back) + (-4.5,0)$) to[short,-o] ($(I2back) + (-3.13,0)$);
\draw ($(I2back) + (-3.08,0)$) to[short] ($(I2back) + (-1.05,0)$);
\draw ($(I2back) + (-4.5,0)$) to ($(I2) + (-4.5,-0.9)$);
\draw ($(I2) + (-4.5,-0.15)$) to ($(I2) + (-4.5,0)$);
\draw ($(I2) + (-4.5,0)$) to[/tikz/circuitikz/bipoles/length=20pt,R,l=$R$,-] ($(I2)+ (-3,0)$);
\draw ($(I2) + (-3,0)$) to[/tikz/circuitikz/bipoles/length=33pt, V, l^=$\widetilde{v}_{\textrm{N,T},2}$, -] ($(I2) + (-1.5,0)$);
\draw ($(I2) + (-3.15,0)$) to[short, o-] ($(I2) + (-3,0)$);
\draw ($(I2) + (-1.5,0)$) to[short] ($(I2) + (-1.05,0)$);
\draw ($(IM) + (-4.5,-0.15)$) to[/tikz/circuitikz/bipoles/length=33pt, V, l_=$v_{\textrm{G},N}$] ($(IM) + (-4.5,-0.9)$);
\draw ($(IMback) + (-4.5,0)$) to[short,-o] ($(IMback) + (-3.13,0)$);
\draw ($(IMback) + (-3.08,0)$) to[short,-o] ($(IMback) + (-1.0,0)$);
\draw ($(IMback) + (-4.5,0)$) to ($(IM) + (-4.5,-0.9)$);
\draw ($(IM) + (-4.5,-0.15)$) to ($(IM) + (-4.5,0)$);
\draw ($(IM) + (-4.5,0)$) to[/tikz/circuitikz/bipoles/length=20pt,R,l=$R$,-] ($(IM)+ (-3.2,0)$);
\draw ($(IM) + (-3,0)$) to[/tikz/circuitikz/bipoles/length=33pt, V, l^=$\widetilde{v}_{\textrm{N,T},N}$, -] ($(IM) + (-1.5,0)$);
\draw ($(IM) + (-3.15,0)$) to[short, o-] ($(IM) + (-3,0)$);
\draw ($(IM) + (-1.5,0)$) to[short] ($(IM) + (-1.05,0)$);
\draw ($(Ir1) + (3,-0.5)$) to[/tikz/circuitikz/bipoles/length=33pt, V, l_=$\widetilde{v}_{\textrm{N,R},1}$, o-o] ($(Ir1) + (1,-0.5)$);
\draw ($(Ir1) + (3.04,-0.5)$) to ($(Ir1) + (4,-0.5)$);
\draw ($(Ir1) + (4,-0.5)$)  to[/tikz/circuitikz/bipoles/length=20pt,R, l_=\mbox{{$R_{{\textrm{in}}}$}}, i>^=\mbox{{$i_{R_{\textrm{in}}}$}}] ($(Ir1) + (4,-1.5)$);
\draw ($(Ir1) + (1,-1.5)$) to[short,o-o] ($(Ir1) + (3,-1.5)$);
\draw ($(Ir1) + (3.04,-1.5)$) to[short] ($(Ir1) + (4,-1.5)$);
\draw ($(Ir2) + (3,-0.5)$) to[/tikz/circuitikz/bipoles/length=33pt, V, o-o] ($(Ir2) + (1,-0.5)$);
\draw ($(Ir2) + (3.04,-0.5)$) to ($(Ir2) + (4,-0.5)$);
\draw ($(Ir2) + (4,-0.5)$)  to[/tikz/circuitikz/bipoles/length=20pt,R, l_=\mbox{{$R_{{\textrm{in}}}$}}, i>^=\mbox{{$i_{R_{{\textrm{in}}}}$}}] ($(Ir2) + (4,-1.5)$);
\draw ($(Ir2) + (1,-1.5)$) to[short,o-o] ($(Ir2) + (3,-1.5)$);
\draw ($(Ir2) + (3.04,-1.5)$) to[short] ($(Ir2) + (4,-1.5)$);
\draw ($(IrM) + (3,0.3)$) to[/tikz/circuitikz/bipoles/length=33pt, V, l_=$\widetilde{v}_{\textrm{N,R},M}$, o-o] ($(IrM) + (1,0.3)$);
\draw ($(IrM) + (3.04,0.3)$) to ($(IrM) + (4,0.3)$);
\draw ($(IrM) + (4,0.3)$)  to[/tikz/circuitikz/bipoles/length=20pt,R, l_=\mbox{{$R_{{\textrm{in}}}$}}, i>^=\mbox{{$i_{R_{{\textrm{in}}}}$}}] ($(IrM) + (4,-0.7)$);
\draw ($(IrM) + (1,-0.7)$) to[short,o-o] ($(IrM) + (3,-0.7)$);
\draw ($(IrM) + (3.04,-0.7)$) to[short] ($(IrM) + (4,-0.7)$);
\draw ($(Ir1) + (6.5,-1.5)$) to[short] ($(Ir1) + (4,-1.5)$);
\draw ($(Ir2) + (6.5,-1.5)$) to[short] ($(Ir2) + (4,-1.5)$);
\draw ($(IrM) + (6.5,-0.7)$) to[short] ($(IrM) + (4,-0.7)$);
\draw  ($(Ir1) + (6.5,-0.5)$) to[/tikz/circuitikz/bipoles/length=30pt,cV, label=\mbox{}] ($(Ir1) + (6.5,-1.5)$);
\draw ($(Ir1) + (9,-0.5)$) to[/tikz/circuitikz/bipoles/length=33pt, V, l_=$v_{\textrm{N},1}$, o-] ($(Ir1) + (6.5,-0.5)$);
\draw  ($(Ir1) + (6.5,-1.5)$) to[short,-o] ($(Ir1) + (9,-1.5)$);
\draw ($(Ir1) + (9,-0.65)$) to [open,v={\small }] ($(Ir1) + (9,-1.4)$);
\draw  ($(Ir2) + (6.5,-0.5)$) to[/tikz/circuitikz/bipoles/length=30pt,cV, label=\mbox{}] ($(Ir2) + (6.5,-1.5)$);
\draw ($(Ir2) + (9,-0.5)$) to[/tikz/circuitikz/bipoles/length=33pt, V, l_=$v_{\textrm{N},2}$, o-] ($(Ir2) + (6.5,-0.5)$);
\draw  ($(Ir2) + (6.5,-1.5)$) to[short,-o] ($(Ir2) + (9,-1.5)$);
\draw ($(Ir2) + (9,-0.65)$) to [open,v={\small }] ($(Ir2) + (9,-1.4)$);
\draw  ($(IrM) + (6.5,0.3)$) to[/tikz/circuitikz/bipoles/length=30pt,cV, label=\mbox{}] ($(IrM) + (6.5,-0.7)$);
\draw ($(IrM) + (9,0.3)$) to[/tikz/circuitikz/bipoles/length=33pt, V, l_=$v_{\textrm{N},M}$, o-] ($(IrM) + (6.5,0.3)$);
\draw  ($(IrM) + (6.5,-0.7)$) to[short,-o] ($(IrM) + (9,-0.7)$);
\draw ($(IrM) + (9,0.15)$) to [open,v={\small }] ($(IrM) + (9,-0.6)$);
\node[] at (-2.5,-0.7) {{\textbf{$\vdots$}}};
\node[] at (3.5,-0.7) {{\textbf{$\vdots$}}};
\node[] at (8.75,-0.7) {{\textbf{$\vdots$}}};
\node[] at (-2.4,2.85) {{$v_{\text{T},1}~~$}};
\node[] at (-2.4,0.85) {{$v_{\text{T},2}~~$}};
\node[] at (-2.4,-2.85) {{$v_{\text{T},N}~~$}};
\node[] at (2.4,2.35) {{$~~v_{\text{R},1}$}};
\node[] at (2.4,0.35) {{$~~v_{\text{R},2}$}};
\node[] at (2.4,-2.65) {{$~~v_{\text{R},M}$}};
\node[] at (-4.55,2.85) {{$v^\prime_{\text{T},1}~~$}};
\node[] at (-4.55,-2.85) {{$v^\prime_{\text{T},N}$}};
\node[] at (-4.55,0.85) {{$v^\prime_{\text{T},2}$}};
\node[] at ($(Ir2) + (2,0.15)$) {{$\widetilde{v}_{\textrm{N,R},2}$}};
\node[] at ($(Ir1) + (9.7,-1)$) {{$v_{\textrm{L},1}$}};
\node[] at ($(Ir2) + (9.7,-1)$) {{$v_{\textrm{L},2}$}};
\node[] at ($(IrM) + (9.7,-0.2)$) {{$v_{\textrm{L},M}$}};
\node[] at ($(Ir1) + (5.65,-1)$) {{$v_{\textrm{LNA}}$}};
\node[] at ($(Ir2) + (5.65,-1)$) {{$v_{\textrm{LNA}}$}};
\node[] at ($(IrM) + (5.65,-0.2)$) {{$v_{\textrm{LNA}}$}};
\draw [dashed] ($(Ir1) + (-3,-7)$) to ($(Ir1) + (-3,-8.5)$);
\draw [dashed] ($(Ir1) + (1,-6.5)$) to ($(Ir1) + (1,-8.5)$);
\draw [dashed] ($(Ir1) + (3,-6.5)$) to ($(Ir1) + (3,-8.5)$);
\node[] at ($(Ir1) + (-4.7,-7.5)$) {$\small{\textrm{Tx signal generator}}$};
\node[] at ($(Ir1) + (-1,-7.5)$) {$\small{\textrm{antenna mutual coupling}}$};
\node[] at ($(Ir1) + (2,-7.5)$) {$\small{\textrm{Rx extrinsic}}$};
\node[] at ($(Ir1) + (2,-8)$) {$\small{\textrm{noise}}$};
\node[] at ($(Ir1) + (6,-7.5)$) {$\small{\textrm{Rx LNA}}$};
\end{circuitikz}
\caption{Linear multiport model of a radio MIMO communication system showing signal generation, antenna mutual coupling, and noise from both extrinsic (i.e., picked up by the antennas) and intrinsic (i.e., generated by low-noise amplifiers and local circuitry) origins. The dependence on the frequency argument ($f$) was dropped to lighten the notation in the figure.}
\label{fig:MIMO-communication-circuit}
\vspace{-0.5cm}
\end{figure*}

\noindent In Section \ref{sec:mutual-self-impedance-computation}, we will resort to this model to derive the off-diagonal elements of the impedance matrices $\bm{Z}_{\text{T}}(f)$ and $\bm{Z}_{\text{R}}(f)$ of the transmit and receive antenna arrays. This is to be opposed to the widely adopted assumption in the antenna design and the analysis of MIMO communication systems where antennas are assumed to be decoupled owing to the half-wavelength spacing, i.e., the impedance matrices $\bm{Z}_{\text{T}}(f)$ and $\bm{Z}_{\text{R}}(f)$ are diagonal matrices.

\section{MIMO communication system model}\label{sec:mimo-system}
\subsection{A circuit-theoretic MIMO communication model}\label{subsec:mimo-model}
We consider the circuit model depicted in Fig. \ref{fig:MIMO-communication-circuit} to study a MIMO communication system where the transmit and receive antennas are at a FF separation distance. There, the joint impedance matrix $\bm{Z}_\text{MIMO}$ models the overall system impedance depending on whether the mutual coupling effects within the transmit and receive array elements are considered or ignored.

\noindent In Fig.~\ref{fig:MIMO-communication-circuit}, the $N$ transmit voltage sources are represented by non-ideal voltage generators $\bm{v}_{\textrm{G}}(f)=\big[ v_{\text{G},1}(f), v_{\text{G},2}(f), \dots, v_{\text{G},N}(f)\big]^\top$, i.e., with internal resistance $R$. Their terminals are connected to the transmitting antennas with the current-voltage pairs $(\bm{v}_\textrm{T}(f),\bm{i}_\textrm{T}(f))$ through the noise voltage sources $\widetilde{\bm{v}}_{\textrm{N,T}}(f)= \big[ \widetilde{v}_{\text{N,T},1}(f), \widetilde{v}_{\text{N,T},2}(f), \dots, \widetilde{v}_{\text{N,T},N}(f)\big]^\top$. Likewise, the receive antenna terminals with the current-voltage pair $(\bm{v}_\textrm{R}(f),\bm{i}_\textrm{R}(f))$ are connected to the low-noise amplifiers (LNAs) through the noise voltage sources $\widetilde{\bm{v}}_{\textrm{N,R}}(f)= \big[ \widetilde{v}_{\text{N,R},1}(f), \widetilde{v}_{\text{N,R},2}(f), \dots, \widetilde{v}_{\text{N,R},M}(f)\big]^\top$. Finally, the LNA is connected to the outside world through the output voltage port $\bm{v}_\textrm{L}(f) = \big[ v_{\text{L},1}(f), v_{\text{L},2}(f), \dots, v_{\text{L},M}(f)\big]^\top$ to which a load impedance of a device can be connected. Since the circuit model in Fig.~\ref{fig:MIMO-communication-circuit} accounts for the extrinsic noise in the transmit/receive antennas and LNAs for more realistic scenarios, it is necessary to specify the statistical properties of all noise voltage sources as we describe hereafter.\vspace{-0.2cm}

\subsection{The noise properties of the circuit model}
\subsubsection{Background noise of antennas}
The multiple-port network $\bm{Z}_\text{MIMO}$ is only composed of passive components which have the same absolute temperature $T$ of the environment \cite{ivrlavc2010toward}. Therefore, the noise of the joint impedance matrix $\bm{Z}_\text{MIMO}$ originates solely from the thermal agitation of the electrons flowing inside its all passive components, a.k.a, thermal noise at the equilibrium temperature $T$ \cite{nyquist1928thermal}. In Fig.~\ref{fig:MIMO-communication-circuit}, the transmit and receive noise voltages $\widetilde{\bm{v}}_{\textrm{N,T}}(f)$ and $\widetilde{\bm{v}}_{\textrm{N,R}}(f)$ model the background noise of transmit/receive antennas. When MC is taken into account within the transmit/receive arrays, the correlations of the $n$th transmit noise voltage $\widetilde{v}_{\textrm{N,T},n}(f)$ and the $m$th receive noise voltage $\widetilde{v}_{\textrm{N,R},m}(f)$ are determined using the truncated Fourier transform \cite{nyquist1928thermal} as follows:

\begin{subequations}
\label{eq:noise-voltage-auto-correlation}
    \begin{align}
        &\lim_{T_0\to\infty}\frac{1}{T_0}\,\mathbb{E}\big[\widetilde{v}_{\textrm{N,T},n}(f)\,\widetilde{v}^*_{\textrm{N,T},n^\prime}(f)\big] ~~=~ 4\,k_{\textrm{b}}\,T\,\Re{\{Z_{\textrm{T},nn^\prime}\}}\nonumber\\
        &\hspace{4cm}\quad ~~\forall \,n,\,n^\prime \in [1,\dots,N],\label{eq:noise-voltage-auto-correlation-Tx}\\
        &\lim_{T_0\to\infty}\frac{1}{T_0}\,\mathbb{E}\big[\widetilde{v}_{\textrm{N,R},m}(f)\,\widetilde{v}^*_{\textrm{N,R},m^\prime}(f)\big] ~=~ 4\,k_{\textrm{b}}\,T\,\Re{\{Z_{\textrm{R},mm^\prime}\}}\nonumber\\
        &\hspace{4cm}\quad \forall \,m,\,m^\prime \in [1,\dots,M].\label{eq:noise-voltage-auto-correlation-Rx}
    \end{align}
\end{subequations}

\noindent For FF communications, when the MC is ignored between the transmit and receive antennas, the transmit/receive noise voltages, $\widetilde{v}_{\textrm{N,T},n}(f)$ and $\widetilde{v}_{\textrm{N,R},m}(f)$, are uncorrelated from each other for all $n \in \{1,\dots, N\}$ and $m \in \{1,\dots, M\}$, which implies that their cross-correlation is zero, i.e.:
\begin{equation}\label{eq:cross-correlation-zero}
\mathbb{E}\big[\widetilde{v}_{\textrm{N,T},n}(f)~ \widetilde{v}_{\textrm{N,R},m}^*(f)\big]  = 0\quad\forall\, n,m.  
\end{equation}
Moreover, all transmit noise voltages $\widetilde{v}_{\textrm{N,T},n}(f)$ can be ignored for FF communication because they are dominated by the transmit power that is injected in the overall transmit generator voltage $\bm{v}_{\textrm{N,T}}(f)$. This is to be opposed to NF communication scenarios where the transmit/receive noise voltages $\widetilde{v}_{\textrm{N,T},n}(f)$ and $\widetilde{v}_{\textrm{N,R},m}(f)$ are correlated, hence the importance of accounting for the transmit voltages in the circuit analysis. 

\subsubsection{The receive LNA model} The LNA is modeled as a noisy frequency flat device with gain $\beta$ such that
\begin{equation}\label{eq:LNA-voltage-gain}
    \bm{v}_{\rm{LNA}}(f) = \beta \,  \,\bm{v}_{\rm{R}_{\textrm{in}}}(f)\,\,[{\textrm{V}}],
\end{equation}
where $\bm{v}_{\rm{R}_{\textrm{in}}}(f)\triangleq - R_{\text{in}}\,\bm{i}_{\textrm{R}}(f)$. For the $m$th amplifier with input impedance $R_{\textrm{in}}$ and noise figure $N_\textrm{f}$, the second-order statistics of the noise voltage, $\widetilde{v}_{\text{N,LNA},m}(f)$, generated inside the LNA are determined using the truncated Fourier transform:
\begin{equation}\label{eq:noise-autocorrelation-amplifier}
    \lim_{T_0\to\infty}\frac{1}{T_0} \,\mathbb{E}[|\widetilde{v}_{\text{N},\text{LNA},m}(f)|^2] = 4\,k_\text{b}\,T\,R_{\textrm{in}}\,(N_\text{f} - 1).
\end{equation}

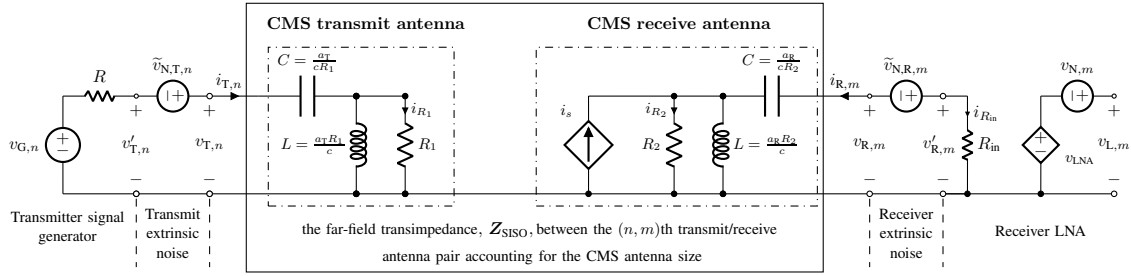
\begin{figure*}[t]
\centering
 \begin{circuitikz}[american voltages, american currents, scale=0.65, every node/.style={transform shape}]
\draw (-4.5,2) to[/tikz/circuitikz/bipoles/length=33pt, V, l_=$v_{\textrm{G},n}$] (-4.5,0);
\draw (-4.5,2) to[/tikz/circuitikz/bipoles/length=20pt,R,l=$R$] (-3,2);
\draw (-4.5,0) to[short,-o] (-3,0);
\draw  (-1.5, 2) to[/tikz/circuitikz/bipoles/length=30pt,V, label=\mbox{}, l_=$\widetilde{v}_{\textrm{N,T},n}$, o-o] (-3, 2);
\draw (-3,0) to[short, o-o] (-1.5,0);
\draw (-3, 2) to [open,v=$v_{\textrm{T},n}^\prime$] (-3,0);
\draw (-1.5,0) to[short,o-*] (1.5,0);
\draw (1.5,0) to[short,*-*] (2.5,0);
\draw (1.5,2) to[short,*-*] (2.5,2);
\draw (-1.5,2) to [short,i>^=$\hspace{-0.2cm}i_{\textrm{T},n}$, o-] (-0.5,2);
\draw (-0.5,2) to[C, label=\mbox{\small{$C=\frac{a_{\text{T}}}{c R_1}$}}, -*] (1.5,2);
\draw (1.5,2)  to[L, l_=\mbox{\small{$L=\frac{a_{\text{T}} R_1}{c}$}}, *-*] (1.5,0);
\draw (2.5,2) to[/tikz/circuitikz/bipoles/length=30pt,R, i>^=\mbox{\small{$i_{R_1}$}}, l=\mbox{\small{$R_1$}}, *-*] (2.5,0);
\draw (-1.5, 2) to [open,v=$v_{\textrm{T},n}$] (-1.5,0)
(2.5,0) to[short, *-*] (6.25,0)
(6.25,0) to[cI, label=\mbox{}] (6.25,2)
(6.25, 2) to[short,-*] (8,2)
(6.25, 0) to[short,-*] (8,0)
(8,2)  to[/tikz/circuitikz/bipoles/length=30pt,R, l_=\mbox{\small{$R_2$}}, i>_=\mbox{\small{$i_{R_2}$}}, *-*] (8,0)
(8, 2) to[short,-*] (9,2)
(8, 0) to[short,-*] (9,0)
(9,0)  to[L, l_=\mbox{\small{$L=\frac{a_{\text{R}} R_2}{c}$}}, *-*] (9,2)
(9,0) to[short,-] (11,0)
(9,2) to[C, label=\mbox{\small{$C=\frac{a_{\text{R}}}{c R_2}$}}, ] (11,2)
(12, 2) to [open,v=$v_{\textrm{R},m}$] (12,0)
(12,2) to [short,i>_=$i_{\textrm{R},m}$] (11,2)
(12,0) -- (11,0)
(13.5, 2) to [open,v=$\hspace{-0.15cm}v_{\textrm{R},m}^\prime$] (13.5,0);
\draw  (13.5,2) to[/tikz/circuitikz/bipoles/length=30pt,V, label=\mbox{}, l_=$\widetilde{v}_{\textrm{N,R},m}$, -o] (12,2);
\draw (13.5,2) to[short, o-] (14,2);
\draw (14,2)  to[/tikz/circuitikz/bipoles/length=20pt,R, l^=\mbox{{$R_{{\textrm{in}}}$}}, i>^=\mbox{{$i_{R_{{\textrm{in}}}}$}}, -*] (14,0);
\draw (12,0) to[short,o-*] (14,0);
\draw (13.5,0) to[short,o-*] (15.5,0);
\draw  (15.5, 2) to[/tikz/circuitikz/bipoles/length=30pt,cV, label=\mbox{}] (15.5, 0);
\draw  (17,2) to[/tikz/circuitikz/bipoles/length=30pt,V, label=\mbox{},l_=$v_{\textrm{N},m}$, o-] (15.5,2);
\draw (15.5,0) to[short,-o] (17,0);
\draw (17,2) to [open,v={{$v_{\textrm{L},m}$} }] (17,0);
\node[] at (5.8,1.7) {\small{$i_{s}$}};
\node[] at (16.3,0.7) {\small{$v_{\text{LNA}}$}};
\node[] at (1.7,3.5) {$\mathbf{CMS ~transmit~ antenna}$};
\node[] at (8.1,3.5) {$\mathbf{CMS ~receive~ antenna}$};
\draw[dashdotted] (-0.3,-0.25) rectangle +(3.8,3.25);
\draw[dashdotted] (5.18,-0.25) rectangle +(5.75,3.25);
\draw (-0.75,-1.6) rectangle +(11.8,5.5);
\draw [dashed] (-1.5, -0.1) to (-1.5, -1.5);
\draw [dashed] (12, -0.1) to (12, -1.5);
\draw [dashed] (13.5, -0.1) to (13.5, -1.5);
\draw [dashed] (-3, -0.1) to (-3, -1.5);
\node[] at (-4.4,-0.5) {$\small{\textrm{Transmitter signal}}$};
\node[] at (-4.4,-0.9) {$\small{\textrm{generator}}$};
\node[] at (5.25,-0.75) {$\small{\text{the far-field transimpedance},\,\bm{Z}_\text{SISO}, \textrm{between the $(n,m)$th transmit/receive}}$};
\node[] at (5.25,-1.3) {$\small{\textrm{ antenna pair accounting for the CMS antenna size}}$};
\node[] at (12.75,-0.4) {$\small{\textrm{Receiver}}$};
\node[] at (12.75,-0.8) {$\small{\textrm{extrinsic}}$};
\node[] at (12.75,-1.2) {$\small{\textrm{noise}}$};
\node[] at (15.5,-0.75) {$\small{\textrm{Receiver LNA}}$};
\node[] at (-2.25,-0.4) {$\small{\textrm{Transmit}}$};
\node[] at (-2.25,-0.8) {$\small{\textrm{extrinsic}}$};
\node[] at (-2.25,-1.2) {$\small{\textrm{noise}}$};
\end{circuitikz}
\caption{Far-field SISO communication model between the $n$th transmit antenna and the $m$th receive antenna where the channel $\bm{Z}_{\text{SISO}}$ consists of transmit/receive electrical CMS's Chu antennas from Fig.~\ref{fig:tm1} with resistance $R_1$ and $R_2$, respectively. Only the receive antenna includes a controlled current source $i_{s}$ by $i_{R_1}$ which models the electromagnetic influence of the transmitter on the receiver in the FF region.}
\label{fig:siso-far-field-system-model}
\vspace{-0.3cm}
\end{figure*}

Additionally, at the $m$th receive antenna, the LNA noise voltage, $\widetilde{v}_{\text{N},\text{LNA},m}(f)$, is uncorrelated with the receive noise voltage, $\widetilde{v}_{\text{N,R},m}(f)$, as well as with the noise voltage of the $n$th transmitter, $\widetilde{v}_{\rm{N,T},n}(f)$, for all transmit and receive antenna pairs ($n,m$). That is to say:
\begin{subequations}\label{eq:noise-cross-correlation-amplifier}
    \begin{align}
    \lim_{T_0\to\infty}\frac{1}{T_0}\,\mathbb{E}\big[\widetilde{v}_{\text{N},\text{LNA},m}(f)\,\widetilde{v}^*_{\text{N,T},n}(f)\big] & = 0\quad\forall n,m,\\
    \lim_{T_0\to\infty}\frac{1}{T_0}\,\mathbb{E}\big[\widetilde{v}_{\text{N},\text{LNA},m}(f)\,\widetilde{v}_{\text{N,R},m}^*(f)\big] & = 0\quad\forall m.
    \end{align}
\end{subequations}

\noindent Now that the statistical noise properties are specified, the characterization of the circuit model in Fig.~\ref{fig:MIMO-communication-circuit} requires the specification of the circuit structure of the joint impedance matrix $\bm{Z}_{\text{MIMO}}$.

\subsection{The input-output relationship of the channel model} Using basic circuit analysis, we establish the input-output relationship of the MIMO communication system between the input voltage $\bm{v}_{\textrm{G}}(f)$ and the output voltage $\bm{v}_{\textrm{L}}(f)$ as (see Appendix~\ref{appendix:input-output-mimo-channel}):
\begin{equation}\label{eq:in-out-relationship}
    \bm{v}_{\textrm{L}}(f) = \bm{H}(f)\,\bm{v}_{\mathrm{G}}(f) + \bm{n}(f),
\end{equation}
where
\begin{subequations}\label{eq:channel-noise-expression}
    \begin{align}
    \bm{H}(f) &= \beta\,R_{\textrm{in}}\,\bm{P}\,\bm{Z}_{\mathrm{RT}} \,\bm{Q}, \\
    \bm{n}(f) &= \bm{v}_{\textrm{N}} + \beta\,R_{\textrm{in}}\,\bm{P}\,\Big[\bm{\widetilde{v}}_{\textrm{N,R}} + \bm{Z}_{\textrm{RT}}\,\bm{Q}\,\Big(\bm{Z}_{\textrm{TR}}\,\bm{P}\,\bm{\widetilde{v}}_{\textrm{N,R}}\Big)\Big],\label{eq:noise-mimo-NF}\\
    &\stackrel{(\textrm{FF})}{\approx}\bm{v}_{\textrm{N}} + \beta\,R_{\textrm{in}}\,\bm{P}\, \bm{\widetilde{v}}_{\textrm{N,R}} ,\label{eq:noise-mimo}
    \end{align}
\end{subequations}
with
\begin{subequations}\label{eq:P-and-Q}
    \begin{align}
        \bm{P} &~\triangleq~ \Big(\bm{Z}_{\textrm{R}} + R_{\textrm{in}}\,\mathbf{I}_{M\times M}\Big)^{-1},\\
        \bm{Q} &~\triangleq~ \Big(\bm{Z}_{\textrm{T}} + R\,\mathbf{I}_{N\times N} - \bm{Z}_{\textrm{TR}} \,\bm{P} \,\bm{Z}_{\textrm{RT}}\Big)^{-1}\label{eq:Q-NF}\\
        &\,\stackrel{(\textrm{FF})}{\approx} \Big(\bm{Z}_{\textrm{T}} + R\,\mathbf{I}_{N\times N}\Big)^{-1}.\label{eq:Q}
    \end{align}
\end{subequations}

\noindent Using (\ref{eq:noise-mimo}) and (\ref{eq:noise-voltage-auto-correlation})--(\ref{eq:noise-cross-correlation-amplifier}), the noise correlation matrix is obtained as:
\begin{subequations}\label{eq:noise-correlation}
\begin{align}
    \bm{R}_{\bm{n}} &= 4\,k_b\,T\,R_{\textrm{in}} \nonumber \\
    &\hspace{0.2cm}\times\Bigg[(N_{\textrm{f}}-1)\,\mathbf{I}_{M\times M} + \beta^2\,R_{\textrm{in}}\Big(\bm{P}\,\Re\{{\bm{Z}_{\textrm{R}}}\}\,\bm{P}^{\textsf{H}} \nonumber\\
    & \hspace{1cm}\times(\mathbf{I}_{M\times M} + \bm{Z}_{\textrm{TR}}^{\textsf{H}}\,\bm{Q}^{\textsf{H}}\,\bm{Z}_{\textrm{RT}}^{\textsf{H}}\,\bm{P}^{\textsf{H}})+ \bm{P}\,\bm{Z}_{\textrm{RT}}\,\bm{Q}\,\Big[\bm{Z}_{\textrm{TR}} \nonumber \\
    &\hspace{1.2cm} \times \bm{P}\,\Re{\{\bm{Z}_{\textrm{R}}\}}\,\bm{P}^{\textsf{H}}\,(\mathbf{I}_{M\times M} + \bm{Z}_{\textrm{TR}}^{\textsf{H}}\,\bm{Q}^{\textsf{H}}\,\bm{Z}_{\textrm{RT}}^{\textsf{H}}\,\bm{P}^{\textsf{H}}) \nonumber \\
    & \hspace{1.4cm} \times \Re{\{\bm{Z}_{\textrm{T}}\}}\,\,\bm{Q}^{\textsf{H}}\,\bm{Z}_{\textrm{RT}}^{\textsf{H}}\,\bm{P}^{\textsf{H}}\Big]\Big)\Bigg], \label{eq:noise-correlation-NF}\\
    &\stackrel{(\textrm{FF})}{\approx} 4\,k_b\,T\,R_{\textrm{in}}\,\Bigg[(N_{\textrm{f}}-1)\,\mathbf{I}_{M\times M} + \beta^2\,R_{\textrm{in}}\Big(\bm{P}\,\Re\{{\bm{Z}_{\textrm{R}}}\}\,\bm{P}^{\textsf{H}} \nonumber\\
    &\hspace{2cm}+\bm{P}\,\bm{Z}_{\textrm{RT}}\,\bm{Q}\,\Re{\{\bm{Z}_{\textrm{T}}\}}\,\,\bm{Q}^{\textsf{H}}\,\bm{Z}_{\textrm{RT}}^{\textsf{H}}\,\bm{P}^{\textsf{H}}\Big)\Bigg].\label{eq:noise-correlation-FF}
\end{align}
\end{subequations}
\noindent The obtained input-output relationship for the circuit model shown in Fig. \ref{fig:MIMO-communication-circuit} is valid for both NF and FF communications when the unilateral approximation is not considered, i.e., as done in (\ref{eq:noise-mimo-NF}), (\ref{eq:Q-NF}) and (\ref{eq:noise-correlation-NF}). Because we focus on FF communications in this work, we further simplify these expressions using the unilateral approximation to obtain Eqs. (\ref{eq:noise-mimo}), (\ref{eq:Q}) and (\ref{eq:noise-correlation-FF}), respectively.

\subsection{The circuit model between a transmit/receive antenna pair}

To find the joint impedance matrix $\bm{Z}_{\text{MIMO}}$, it is enough to specify the SISO transimpedance between any transmit/receive antenna pair. Such description is valid because we consider the Chu's CMS model for each antenna element. This model belongs to the family of minimum scattering antennas \cite{kahn1965minimum} which allows one to determine the off-diagonal elements of the impedance matrices $\bm{Z}_{\textrm{T}}$ and $\bm{Z}_{\textrm{R}}$ as well as the transmit-receive transimpedance $\bm{Z}_{\text{RT}}$ by examining the pairwise isolated mutual impedances between two CMS antennas. Indeed, it has been observed in practice that this approach gives accurate predictions of MC between pairs of approximately CMS antennas such as dipoles and helical antennas \cite{andersen1974coupling}.

\noindent Fig.~\ref{fig:siso-far-field-system-model} depicts the SISO communication model when the circuit model of Chu's CMS antennas described in Fig.~\ref{fig:tm1} is adopted in the pairwise communication model from Fig.~\ref{fig:MIMO-communication-circuit} by letting $N=M=1$.

\noindent It is important to notice that the receive CMS antenna model in Fig.~\ref{fig:siso-far-field-system-model} includes an additional controlled current source as opposed to the CMS's Chu antenna model in Fig.~\ref{fig:tm1}. This is because we consider the electromagnetic influence of the transmit antenna on the receive antenna for FF communication only. In other words, omitting a current source from the transmit antenna model corresponds to the circuit-level unilateral approximation which was already described at the impedance level in (\ref{eq:unilateral-approx}).

\noindent To fully characterize the MIMO model in (\ref{eq:in-out-relationship})--(\ref{eq:noise-correlation}), one must find the transmit-receive transimpedance $\bm{Z}_{\text{RT}}(f)$ as well as the transmit and receive impedance matrices $\bm{Z}_{\text{T}}(f)$ and $\bm{Z}_{\text{R}}(f)$. We highlight the fact that the description of the impedance matrices is independent of the input-output relationship pertaining to the circuit model at hand. As one example, one can obtain the impedance matrices from simulations or measurements for a particular antenna array design, then inject them back into (\ref{eq:in-out-relationship})--(\ref{eq:noise-correlation}) to pursue the information-theoretic circuit analysis. In this paper, we only make assumptions about the element size and spacing to derive the impedance matrices entirely through theoretical analysis, i.e., with no assumption regarding the type of antenna elements in use. The analytical expressions of these impedance matrices based on the Chu's CMS antennas will be established in the next section.\vspace{-0.2cm}

\section{Derivation of the impedance matrices}
\label{sec:mutual-self-impedance-computation}
\begin{figure}[h!]
    \hspace{0.3cm}
    \begin{subfigure}{0.45\textwidth}
        \centering
        \includegraphics[scale=0.55]{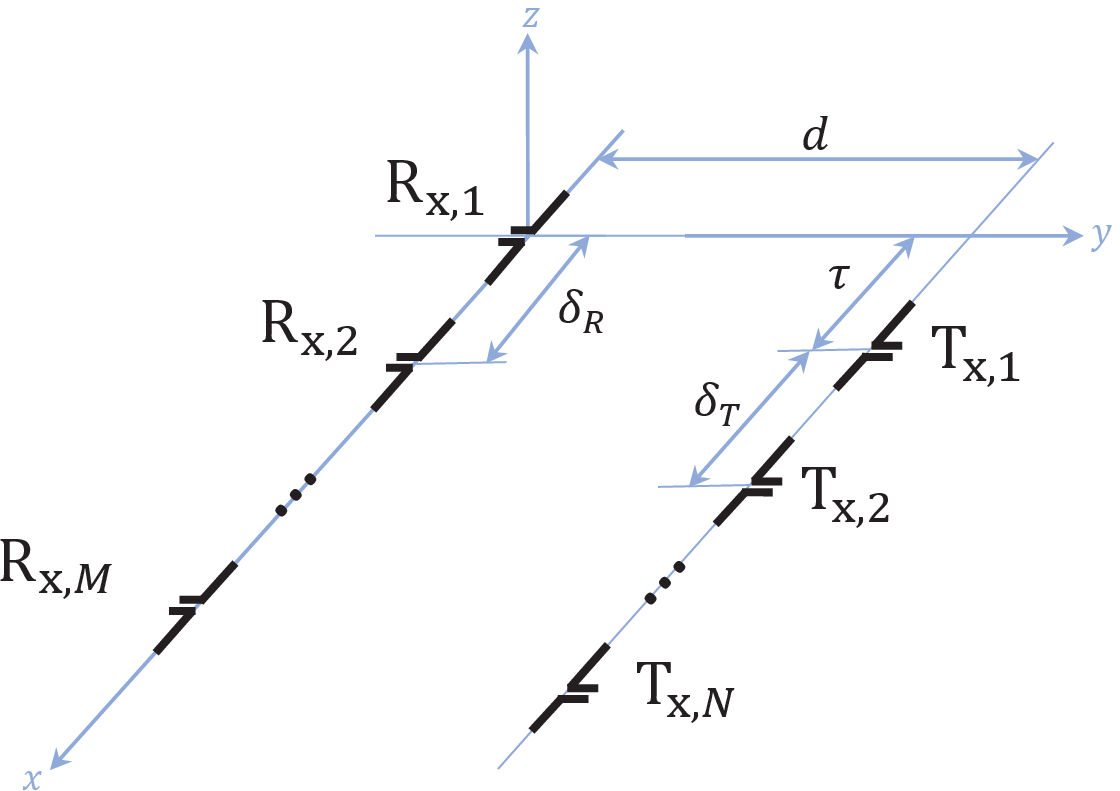}
        \caption{colinear configuration}
        \label{fig:mimo-colinear-config}
    \end{subfigure}
\quad
    \begin{subfigure}{.45\textwidth}
        \centering
        \includegraphics[scale=0.55]{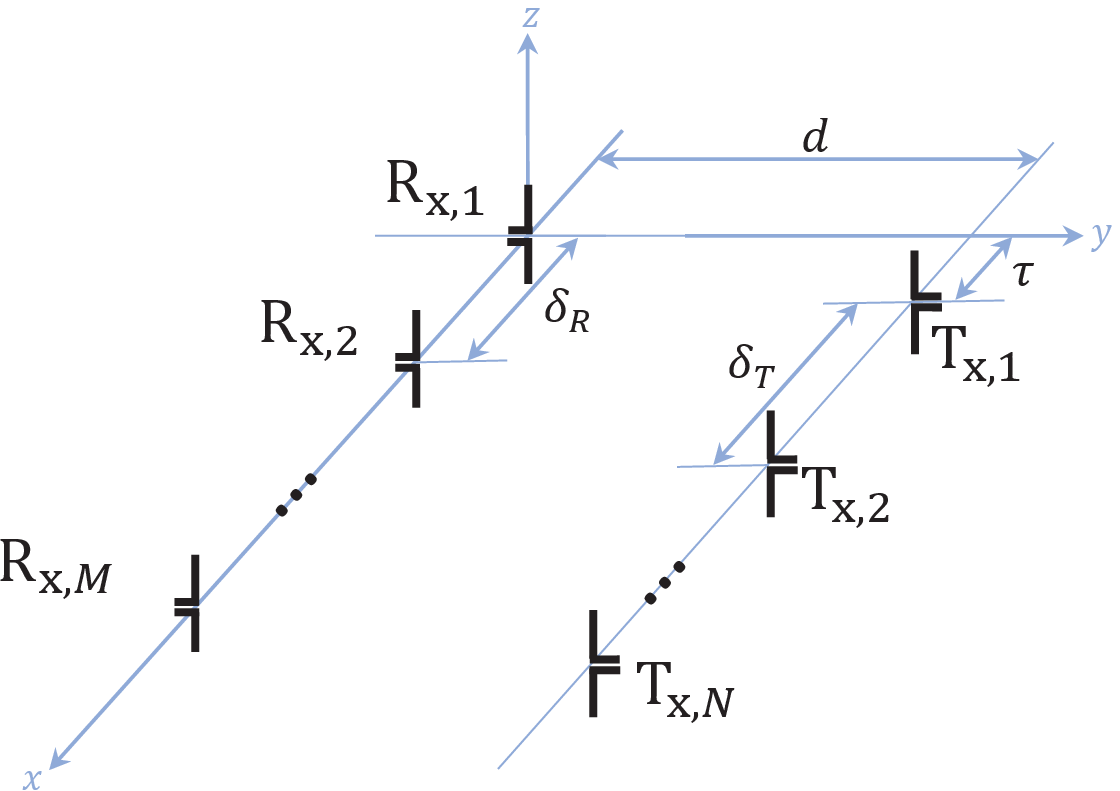}
        \caption{parallel configuration}
        \label{fig:mimo-parallel-config}
    \end{subfigure}
    \vspace{-0.17cm}
\caption{Two special orientations between the transmit and receive ULAs where the axes of the transmit and receive antennas are: (a) colinear to the array axes in the colinear configuration, and (b) perpendicular to the array axes in the parallel configuration.}
\label{fig:mimos-geometry-configurations}
\vspace{-0.8cm}
\end{figure}

In this section, we derive the transimpedance matrix $\bm{Z}_{\textrm{RT}}$ for FF communications using $i)$ line-of-sight (LoS) channels based on Friis' equation and $ii)$ Rayleigh fading channels. We also find the impedance matrices $\bm{Z}_{\textrm{T}}$ and $\bm{Z}_{\textrm{R}}$ based on the MC model in (\ref{eq:mutual-coupling-SISO}) for two standard configurations of uniform linear antenna arrays (ULAs):
\begin{itemize}[leftmargin=*]
    \item \textit{colinear arrays} as depicted in Fig.~\ref{fig:mimo-colinear-config} where the axes of the transmit and receive antennas within the arrays (i.e., inter-element antennas) are colinear to the axes of their respective arrays,
    \item \textit{parallel arrays} as depicted in Fig.~\ref{fig:mimo-parallel-config} where the axes of the transmit and receive antennas within the arrays are perpendicular to the axes of their respective arrays.\vspace{-0.24cm}
\end{itemize}

\subsection{Derivation of the FF transimpedance $\bm{Z}_{\textrm{RT}}$}\label{subsec:FF-mutual-impedance}
To describe the electromagnetic influence of the transmitter on the receiver, the FF transimpedance must account for the propagation properties of the propagation channel. In this paper, we restrict our attention to the following widely encountered MIMO channels:
\begin{itemize}[leftmargin=*]
    \item \textit{line of sight channels (LoS)} governed by the Friis’ equation which leads to rank-1 channels.
    \item \textit{Flat fading Rayleigh channels} with spatial correlation due to MC between the transmit and receive arrays.
\end{itemize}
For LoS channels, we use the Friis' equation to relate the radiated power from transmit antenna $p$ and the received power at receive antenna $q$. In fact, the squared magnitudes of the transmit and the receive currents shown in Fig.~\ref{fig:siso-far-field-system-model} are related as follows:
\begin{equation}\label{eq:friss-current-magnitude}
|i_{s}(f)|^2 = 4\,|i_{R_1}(f)|^2\left(\frac{c}{4{\pi}fd^{\frac{\alpha}{2}}}\right)^2G_\textrm{T}\,G_\text{\textrm{R}}\,\frac{R_1}{R_2}\,\, [\textrm{A}^2].
\end{equation}
In (\ref{eq:friss-current-magnitude}), $\alpha$ is the path-loss exponent and $d$ is the distance between the transmit and receive antennas which have gains $G_{\textrm{T}}$ and $G_{\textrm{R}}$, respectively. For transmit/receive Chu's CMS antennas, we have that $G_{\textrm{T}}=G_{\textrm{R}} = 3/2$ in the equatorial plane \cite[chapter 6]{harrington1961pp}. Using basic circuit analysis, we show that the FF LoS transimpedance matrix $\bm{Z}_{\mathrm{RT}}^{\textrm{LoS}}(f)$ is expressed as (see Appendix \ref{appendix:FF-mutual-impedance}):

\begin{equation}\label{eq:FF-mutual-impedance-line-of-sight-matrix}
    \begin{aligned}[b]
    \bm{Z}_{\mathrm{RT}}^{\textrm{LoS}}(f) &= \frac{c\,\sqrt{G_\text{T}\,G_\text{R}}}{2{\pi}fd^{\frac{\alpha}{2}}}~\textrm{diag}\left(\Re{\{\bm{Z_{\textrm{R}}}(f)\}}\right)^{\frac{1}{2}}\,\bm{a}_\mathrm{R}(\theta_{\mathrm{R}})\,\bm{a}_{\mathrm{T}}^{\mathsf{T}}(\theta_{\mathrm{T}})  \\
    &\hspace{1.2cm}\times\textrm{diag}\left(\Re{\{\bm{Z_{\textrm{T}}}(f)\}}\right)^{\frac{1}{2}}\,e^{-j\phi}~[\Omega].
    \end{aligned}
\end{equation}

\noindent Here, $\theta_{\textrm{T}}$ and $\theta_{\textrm{R}}$ are the angles of arrival and departure defined w.r.t. the broadside axis of the transmit and receive arrays, respectively, and
\begin{subequations}
    \begin{align}
        \phi & = \pi - \textrm{arctan}\left(2\pi f a_{\textrm{T}}/c\right) - \textrm{arctan}\left(2\pi f a_{\textrm{R}}/c\right),\label{eq:relative-phase}\\
        \bm{a}_\mathrm{T}(\theta_{\mathrm{T}})&=\bigg[1, e^{-\textrm{j}2\pi f\delta_{\textrm{T}} \cos(\theta_{\textrm{T}})/c},\hdots, e^{-\textrm{j}2\pi(N-1) f\delta_{\textrm{T}} \cos(\theta_{\textrm{T}})/c} \bigg]^\top,\label{eq:steering-vector-Tx}\\
        \bm{a}_\mathrm{R}(\theta_{\mathrm{R}})&=\bigg[1, e^{-\textrm{j}2\pi f\delta_{\textrm{R}} \cos(\theta_{\textrm{R}})/c},\hdots, e^{-\textrm{j}2\pi(M-1) f\delta_{\textrm{R}} \cos(\theta_{\textrm{R}})/c} \bigg]\label{eq:steering-vector-Rx}.
    \end{align}
\end{subequations}

\noindent To generalize the LoS transimpedance matrix in (\ref{eq:FF-mutual-impedance-line-of-sight-matrix}) to rich-scattering propagation environments under MC, we adopt the correlated Rayleigh fading model which is given by \cite{saab2021optimizing}:
\begin{equation}\label{eq:correlated-Rayleigh-fading}
\begin{aligned}[b]
    \bm{C}(f) &= \textrm{diag}\left(\Re{\{\bm{Z_{\textrm{R}}}(f)\}}\right)^{-\frac{1}{2}}\,\Re{\{\bm{Z_{\textrm{R}}}(f)\}}^{\frac{1}{2}}\\
    &\hspace{1.2cm}\times\bm{F}\,\Re{\{\bm{Z_{\textrm{T}}}(f)\}}^{\frac{1}{2}}~\textrm{diag}\left(\Re{\{\bm{Z_{\textrm{T}}}(f)\}}\right)^{-\frac{1}{2}},
    \end{aligned}
\end{equation}
 
\noindent where each $ij$-entry of the small-scale fading propagation channel $\bm{F}$ is a realization from the complex centered Gaussian distribution with variance $1/2$ per dimension, i.e.,  $\bm{F}_{ij}\sim\mathcal{CN}(0,1/2)$. In (\ref{eq:correlated-Rayleigh-fading}), we characterized the spatial channel correlation entirely based on the impedance matrices $\bm{Z_{\textrm{T}}}(f)$ and $\bm{Z_{\textrm{R}}}(f)$ because it follows the same correlation properties as the isotropic environmental noise which is given by the real part of the impedance matrix \cite{mezghani2020massive}. Finally, we use the correlated fading model $\bm{C}(f)$ in (\ref{eq:correlated-Rayleigh-fading}) to obtain the FF Rayleigh transimpedance matrix as
\begin{equation}\label{eq:FF-mutual-impedance-gaussian}
    \bm{Z}_{\mathrm{RT}}^{\textrm{Rayleigh}}(f) = ~\frac{c}{2{\pi}fd^{\frac{\alpha}{2}}}~\Re{\{\bm{Z_{\textrm{R}}}(f)\}}^{\frac{1}{2}}~\bm{F}~\Re{\{\bm{Z_{\textrm{T}}}(f)\}}^{\frac{1}{2}}~[\Omega].
\end{equation}

\begin{figure*}[bp]
\begin{equation}\label{capacity}
    \begin{aligned}[b]
        C = \max_{\bm{Z}_{\text{MIMO}},\,\mathbb{P}_{\mathbf{\textsf{v}}_{\textrm{T}}}} I(\mathbf{\textsf{v}}_{\textrm{T}}(t);\mathbf{\textsf{v}}_{\textrm{L}}(t))\,\,[{\textrm{bits/s}}] &= \max_{\bm{Z}_{\text{MIMO}},\,\mathbb{P}_{\mathbf{\textsf{v}}_{\textrm{T}}}}~ \sum_{r=0}^{R-1}\int_{0}^{\infty}{\log_2\left(1+ P_r\,\lambda_r\Big[\bm{H}^{\textsf{H}}(f)\,\bm{R}_{\bm{n}}^{-1}\,\bm{H}(f)\Big]\right)\textrm{d}f}\\
        &\approx \,\max_{\mathbb{P}_{\mathbf{\textsf{v}}_{\textrm{T}}}}~ \sum_{r=0}^{R-1}\int_{0}^{\infty}{\log_2\Big(1+ P_r}\,\lambda_r\Big[\bm{H}^{\textsf{H}}(f)\,\bm{R}_{\bm{n}}^{-1}\,\bm{H}(f)\Big]\Big)\bigg|_{\bm{Z}_{\text{MIMO}}=\bm{Z}^{\text{Chu}}_{\text{MIMO}}}\,\,\textrm{d}f.
    \end{aligned}
\end{equation}
\end{figure*}

\subsection{Derivation of the transmit/receive impedances $\bm{Z}_{\textrm{T}}$ and $\bm{Z}_{\textrm{R}}$}\label{subsec:self-impedances-matrices}

Unlike the FF transimpedance matrix derived in Section \ref{subsec:FF-mutual-impedance}, the impedance matrices $\bm{Z}_{\textrm{T}}$ and $\bm{Z}_{\textrm{R}}$ of the transmit and receive antenna arrays should take into account the MC effects induced in the near-field region between the closely spaced antennas within the same array. This is because the significant power differences of the radiated spherical waves in the near-field is a function of the relative orientation between the antenna elements within the array, hence the importance of considering the geometric configuration of the array to characterize the MC effects \cite{akrout2021achievable}.

\noindent  For this purpose, we use the MC model $\mathcal{MC}(d, \beta, \gamma)$ in (\ref{eq:mutual-coupling-SISO}) describing the mutual impedance expression between two closely-spaced Chu's CMS antennas. For each $p$th and $q$th antenna element within the array, this MC model depends on a triplet of parameters ($d_{{pq}}$, $\beta_{{pq}}$, $\gamma_{{pq}}$) as depicted in Fig.~\ref{fig:two-hertz-antennas}. These are the distance $d_{{pq}}$ between the two antennas and their orientation angles $\beta_{pq}$ and $\gamma_{pq}$ w.r.t. their connecting axis. For the colinear and parallel configurations of ULAs illustrated in Fig.~\ref{fig:mimos-geometry-configurations}, the respective triplets are given in Table~\ref{fig:params-MC-model}. Once the MC model is found for each ($p$,$q$) antenna pair within the transmit and receive arrays, the off-diagonal elements of the impedance matrices $\bm{Z}_{\textrm{T}}(f)$ and $\bm{Z}_{\textrm{R}}(f)$ are determined, while their diagonal elements are the self-impedances of Chu's CMS antennas given in (\ref{eq:Z1-chu}). The obtained impedance matrices are fully characterizing the MC effects within the transmit and receive arrays.

\begin{table}[H]
\centering
\caption{Summary of the parameters of the MC model $\mathcal{MC}(d, \beta, \gamma)$ in (\ref{eq:mutual-coupling-SISO}) between an arbitrary antenna pair ($p$,$q$) within the uniform linear array with a uniform spacing of $\delta$ [m].}
\begin{tabular}[t]{ccc}
\toprule
{\textbf{Parameter}}& {\textbf{Colinear antenna pair}} &{\textbf{Parallel antenna pair}} \\ \midrule
{$\beta_{pq}$}   & {0} & {$\pi/2$} \\ \midrule
{$\gamma_{{pq}}$}   & {$\pi$}  & {$\pi/2$} \\ \midrule
{$d_{{pq}}$}   &  {$|p - q| \cdot \delta$} & {$|p - q| \cdot \delta$} \\
\bottomrule
\end{tabular}
\label{fig:params-MC-model}
\end{table}

\section{Achievable rate and optimum inter-antenna spacing optimization}

To find the maximum achievable rate, it is necessary to optimize the mutual information of the overall MIMO system at the disposal of the system designer. The resulting mutual information can be interpreted as the highest of all achievable rates. By inspecting Fig.~\ref{fig:MIMO-communication-circuit}, it is seen that the transmitted waveform  $\mathbf{\textsf{v}}_{\textrm{T}}(t)$ is under the full control of the system designer. Moreover, because the MIMO communication system is physically consistent modeled, it is also possible to partially design the joint impedance matrix $\bm{Z}_\text{MIMO}$ by choosing the appropriate impedances $\bm{Z}_\text{T}$ and $\bm{Z}_\text{R}$ to obtain the optimal transmit and receive antenna structures that maximize the achievable rate. The maximum mutual information is thus given by (\ref{capacity}). There, the last approximation follows from the fact that we consider Chu's CMS antennas which are the ``optimal antennas'' for broadband FF communications under the finite size constraint. Moreover, $I(\mathbf{\textsf{v}}_{\textrm{T}}(t);\mathbf{\textsf{v}}_{\textrm{L}}(t))$ is the mutual information per unit time between the two random processes representing the input and output signals of the communication system~\cite{gel1957computation,gallager1968information}. Additionally, $P_r$ and $\lambda_r$ denote the allocated power and the eigenmode of the $r$th SISO eigenchannel, respectively \cite{heath2018foundations}. Moreover, $\mathbb{P}_{\mathbf{\textsf{v}}_{\textrm{T}}}$ is the probability measure on the space of possible generator voltages, $\mathbf{\textsf{v}}_{\textrm{T}}(t)$, which for any finite set of time instants $\{t_1,t_2,\ldots,t_n\}$ specifies the joint cumulative distribution:
\begin{equation}\label{eq:cdf-generator}
\begin{aligned}[b]
    &\mathbb{P}_{\mathbf{\textsf{v}}_{\textrm{T}}}[\mathbf{\textsf{v}}_{\textrm{T}}(t_1)\leq v_1,\mathbf{\textsf{v}}_{\textrm{T}}(t_2)\leq v_2,\ldots, \mathbf{\textsf{v}}_{\textrm{T}}(t_n)\leq v_n] \\
    &\hspace{4cm}\forall(v_1,v_2,\ldots,v_n)\in \mathbb{R}^n.
\end{aligned}
\end{equation}

\noindent Under a total power budget $P_{\textrm{max}}$ supplied by the generator and optimum power allocation $P_{r}^{\tiny{\starletfill}}(f)$, the achievable rate in (\ref{capacity}) becomes \cite{heath2018foundations}:
\begin{equation}\label{AWGN_Capacity-noise-correlation}
    C = \sum_{r=0}^{R-1}\int_{0}^{\infty}{\log_2\left(1+ P_r^{\tiny{\starletfill}}\,\lambda_r\Big[\bm{H}^{\textsf{H}}(f)\,\bm{R}_{\bm{n}}^{-1}\,\bm{H}(f)\Big]\right)\textrm{d}f}\,\,[{\textrm{bits/s}}],
\end{equation}
with
\begin{equation}\label{eq:power-constraint-mimo}
    P_r^{\tiny{\starletfill}}(f) = \max\left(0,\,\frac{1}{\nu}-\frac{1}{ \lambda_{r}\big[\bm{H}^{\textsf{H}}(f)\,\bm{R}^{-1}_{\bm{n}}\,\bm{H}(f)\big]}\right).
\end{equation}

\noindent Here, $\nu$ must ensure that $\int_{0}^{\infty}\,\sum_{r=0}^{R-1}\,P_r^{\tiny{\starletfill}}(f) \,\textrm{d}f= P_{\textrm{max}}$. The design of the probability law of the generator in (\ref{eq:cdf-generator}) should also satisfy this condition.
 
 \noindent While the optimization of the achievable rate assumes the joint impedance matrix for Chu's CMS antennas, $\bm{Z}_{\text{MIMO}}^{\text{Chu}}$, to be optimal, it is still possible to optimize the latter w.r.t. the spacing-to-antenna-size ratio $\delta/a$. By requiring that both transmit and receive arrays of tightly-coupled massive MIMO to be purely resistive, we show the following result:
 
 \begin{result}[\textbf{condition for tight coupling}]\label{result:condition-tight-coupling}
In the limit of infinite uniform linear Chu's CMS arrays (i.e., $N,M\rightarrow\infty$) with colinear orientation and quasi-continuous aperture (i.e., $\delta/\lambda\rightarrow0$), the optimum spacing-to-antenna-size ratio $(\delta/a)^{\tiny{\starletfill}}$ between Chu's CMS antennas of size $a$ is given by:
\begin{equation}\label{eq:result-1-formula}
    (\delta/a)^{\tiny{\starletfill}} = \sqrt[3]{6 \,\zeta(3)}\,  \approx 1.932,
\end{equation}

\noindent where $\zeta(\cdot)$ is the Riemann zeta function\footnote{This result violates the assumption of the MC model in (\ref{eq:mutual-coupling-SISO}) between two Chu's CMS antennas whose spheres do not overlap (i.e., $\delta/a > 2$), and hence should be interpreted as an approximate result that we validate numerically in Section \ref{sec:results}.}.
\end{result}

\begin{proof}
See Appendix~\ref{appendix:proof-result1}.
\end{proof}

 \noindent Given the condition for tight coupling obtained in Result~\ref{result:condition-tight-coupling} and the expression of the achievable rate in (\ref{capacity}) under the antenna size constraint, we are ready to examine the effect of MC on the overall MIMO system performance.
 \section{Numerical results and discussions}\label{sec:results}
\begin{figure}[h!]
    \begin{subfigure}{0.45\textwidth}
        \vspace{0.4cm}
        \centering
        \includegraphics[scale=0.4]{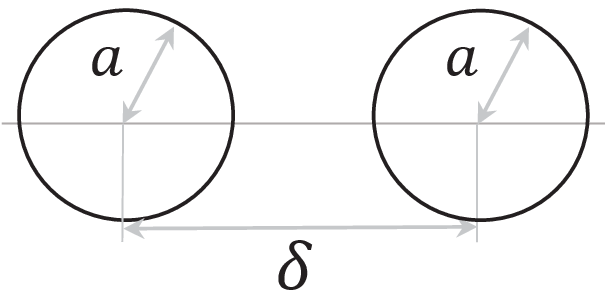}
        \caption{\textit{weakly} coupled antennas}
        \label{fig:non-intersecting-Chu-spheres}
    \end{subfigure}
    \qquad
    \begin{subfigure}{.45\textwidth}
    \vspace{0.1cm}
        \centering
        \includegraphics[scale=0.4]{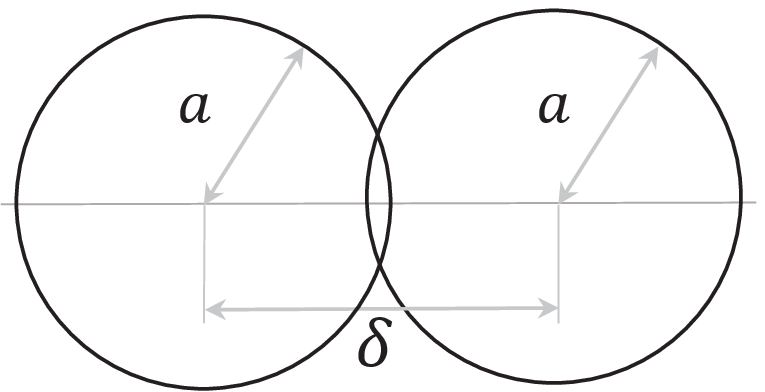}
        \caption{\textit{tightly} coupled antennas}
        \label{fig:overlapping-Chu-spheres}
    \end{subfigure}
\caption{Two possible configurations of the Chu's spheres of neighboring Chu's CMS antennas within transmit/receive arrays: (a) non-overlapping Chu's spheres when $\delta/a > 2$, and (b) overlapping Chu's spheres when $\delta/a < 2$.}
\label{fig:Chu-spheres-configurations}
\end{figure}

In this section, we present simulation results for the maximum achievable rate of FF communications under antenna size constraint and mutual coupling at both the transmitter and the receiver. Since studying the channel is crucial for exploiting the CSI to improve the mutual information  in wireless MIMO transmissions, we start by inspecting the behavior of SIMO/MISO SNRs and achievable rates in (\ref{capacity}) for both the colinear and parallel array configurations shown in Fig.~\ref{fig:mimos-geometry-configurations}. In all simulations, we consider broadband transmit/receive antenna arrays whose elements are Chu's CMS antennas operating between $f_{\textrm{min}}=100\,[\textrm{MHz}]$ and $f_{\textrm{max}}=30\,[\textrm{GHz}]$ with respective wavelengths being denoted as $\lambda_{\textrm{min}}$ and $\lambda_{\textrm{max}}$, respectively. Without loss of generality, we set both the angles of arrival and departure $\theta_{\textrm{T}}$ and $\theta_{\textrm{R}}$ in (\ref{eq:FF-mutual-impedance-line-of-sight-matrix}) to zero, which corresponds to the broadside/frontfire communication scenario. Unless specified otherwise, we set the array spacings to $\delta_\textrm{T}=\delta_\textrm{R}=0.5\,[\textrm{cm}]$, the antenna sizes to $a_\text{T}=a_\text{R} = \delta/\sqrt[3]{6 \,\zeta(3)}$ according to the condition for tight coupling in Result~\ref{result:condition-tight-coupling}. We also set the total power to $P_{\textrm{max}} = 2\,[\textrm{W}]$, the path-loss exponent $\alpha$ to 3.5, and the separation distance between the transmit and receive arrays to $d=30\,\lambda_{\textrm{min}}$ to ensure that the arrays are in their mutual FF regions at any frequency $f\in [f_{\textrm{min}}, f_{\textrm{max}}]$. We also study the achievable rate over three distinct bands $BW_1=[680\,\textrm{MHz}, 720\,\textrm{MHz}]$, $BW_2=[2.45\,\textrm{GHz}, 2.55\,\textrm{GHz}]$ and $BW_3=[24.9\,\textrm{GHz}, 25.1\,\textrm{GHz}]$ just like the carrier aggregation technique used for high-performance 4G/5G networks. Note that we do not use the standard ``central frequency'' terminology as it is only relevant to narrow-band systems. To control the MC effects, we increase or decrease the antenna sizes $a_{\textrm{T}}$ and $a_{\textrm{R}}$ as a function of the fixed spacing $\delta$ as shown in Fig. \ref{fig:Chu-spheres-configurations}.\vspace{-0.2cm}

\subsection{SIMO/MISO achievable rates and spectral efficiencies}
\begin{figure*}[h!]
    \centering
    \hspace{-0.2cm}
    \subcaptionbox{MISO with \textit{tightly} coupled antennas.\label{fig:MISO-coupling}}{{\hspace{-0.5cm}
    \includegraphics[scale=0.3]{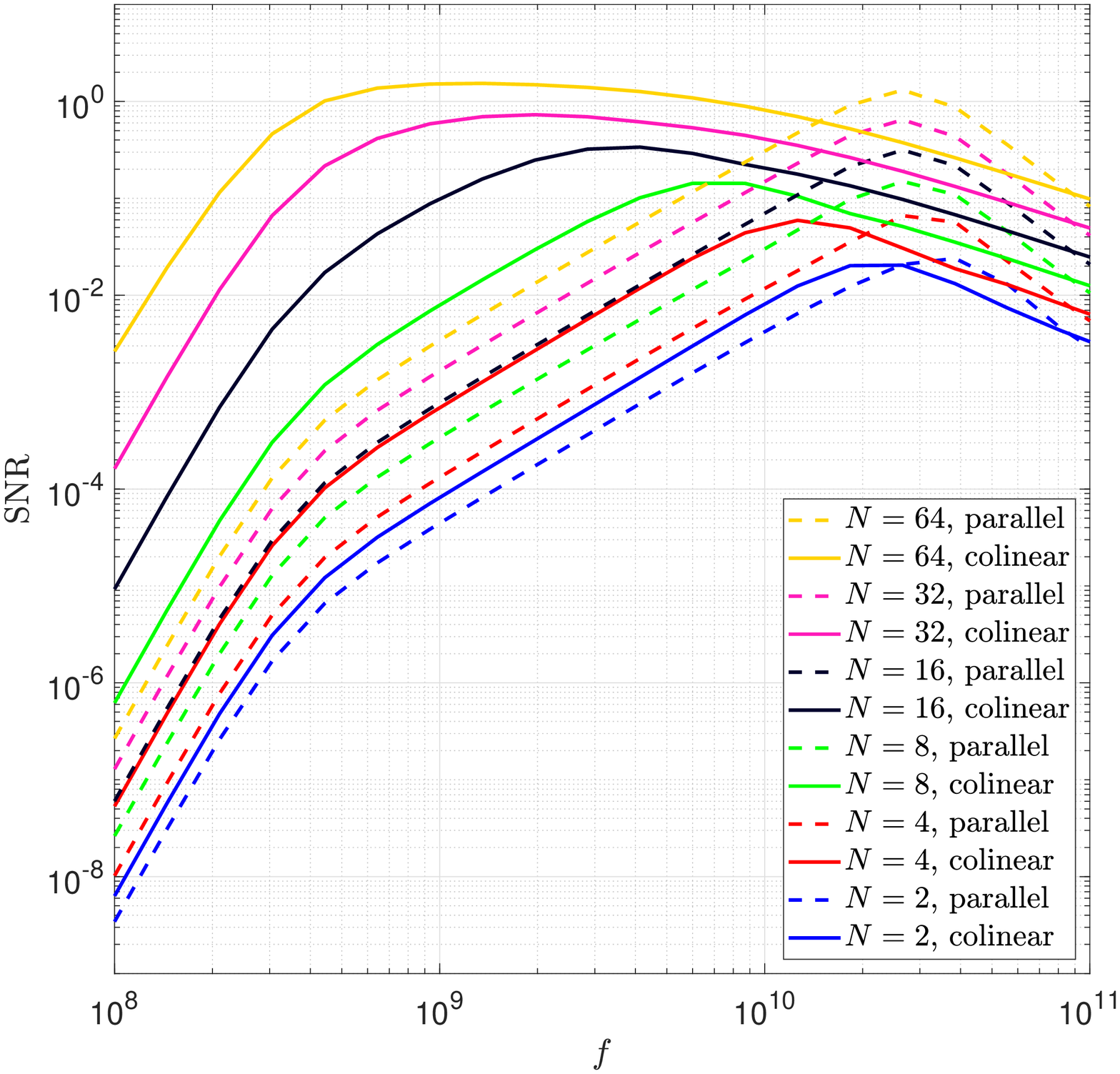}
    }}
    \subcaptionbox{SIMO with \textit{tightly} coupled antennas.\label{fig:SIMO-coupling}}{{
    \includegraphics[scale=0.3]{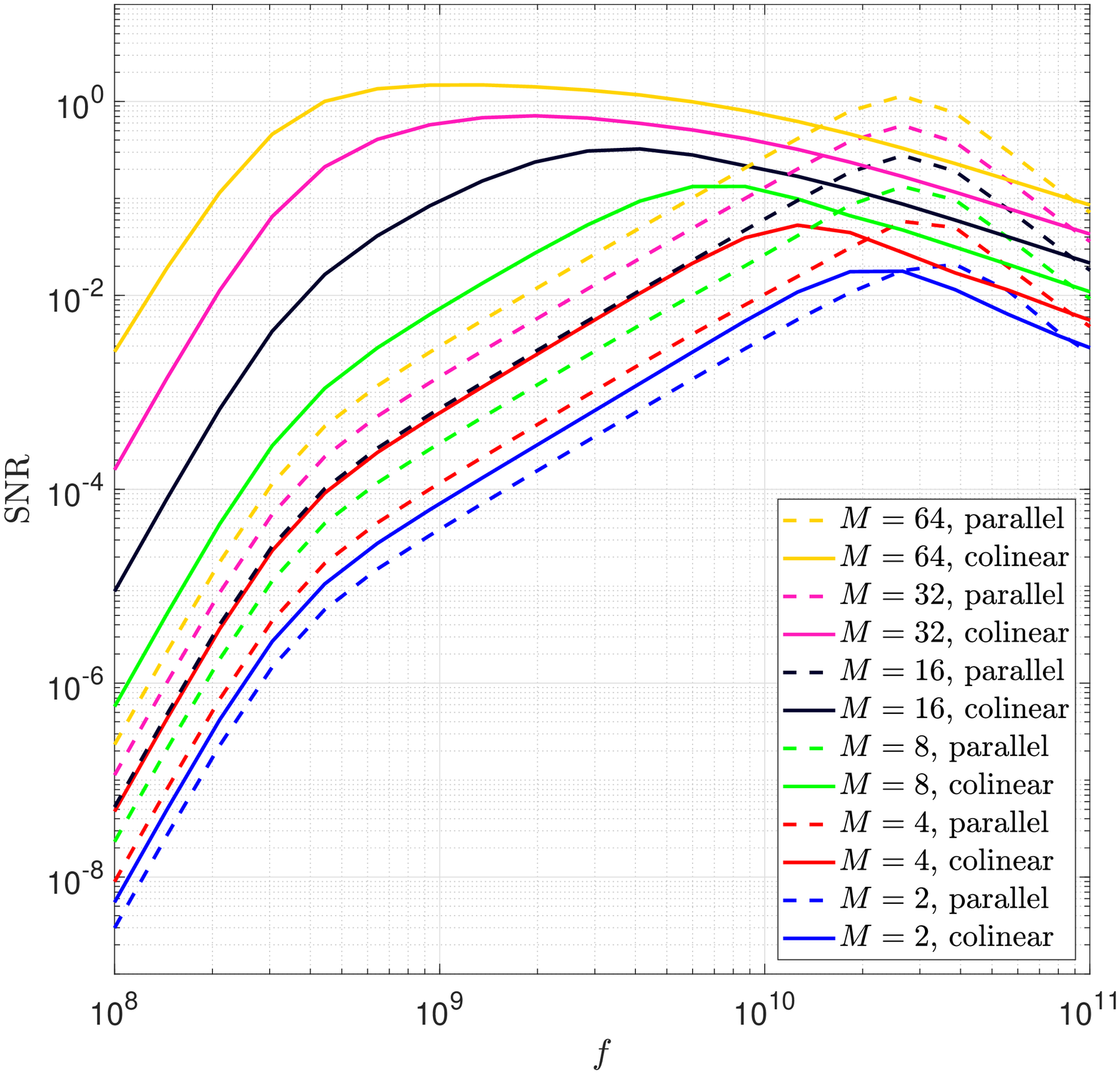}
    }}
    \subcaptionbox{MISO with \textit{weakly} coupled antennas.\label{fig:MISO-no-coupling}}{{\hspace{-0.5cm}
    \includegraphics[scale=0.3]{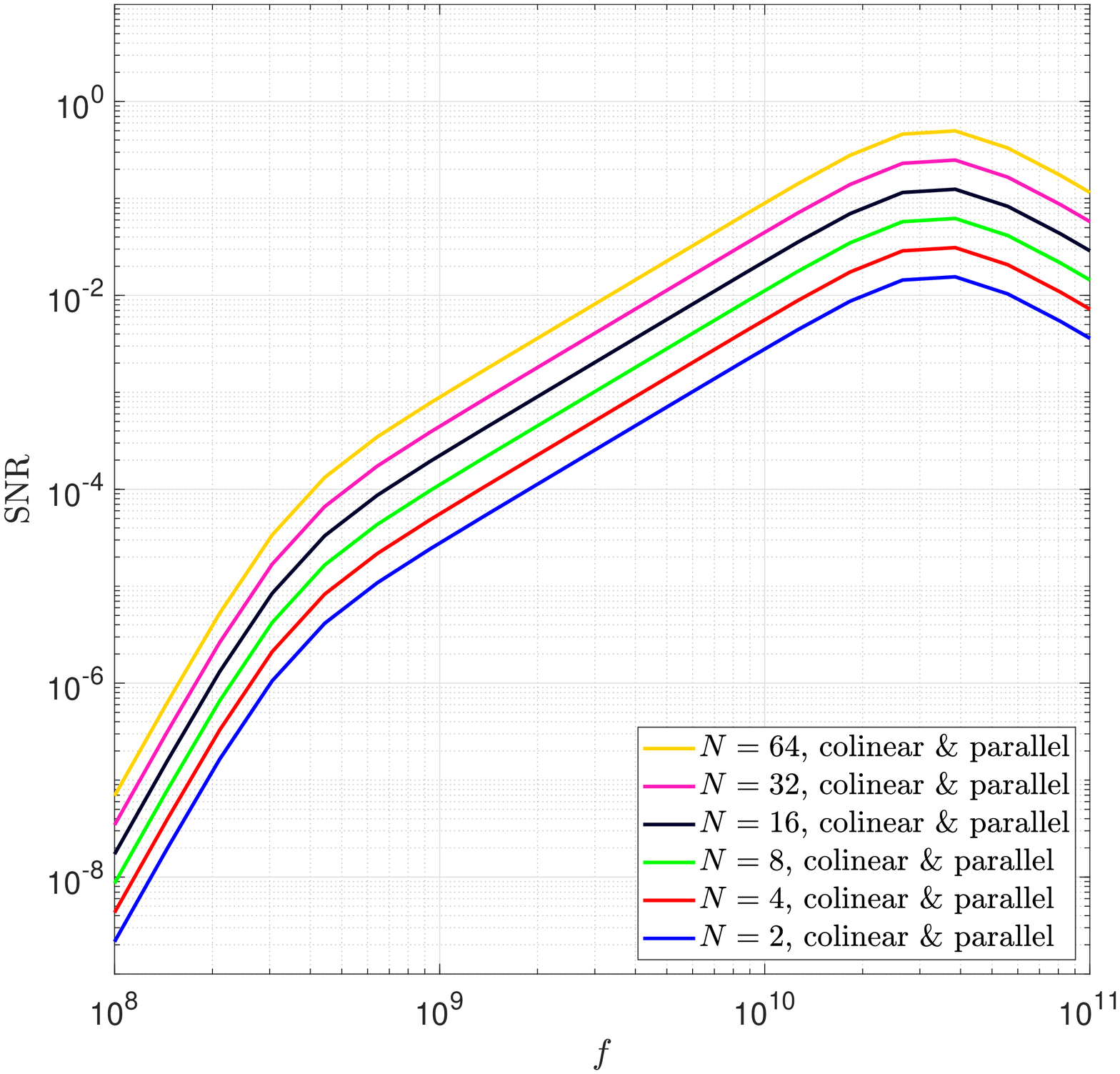}
    }}
    \subcaptionbox{SIMO with \textit{weakly} coupled antennas.\label{fig:SIMO-no-coupling}}{{
    \includegraphics[scale=0.3]{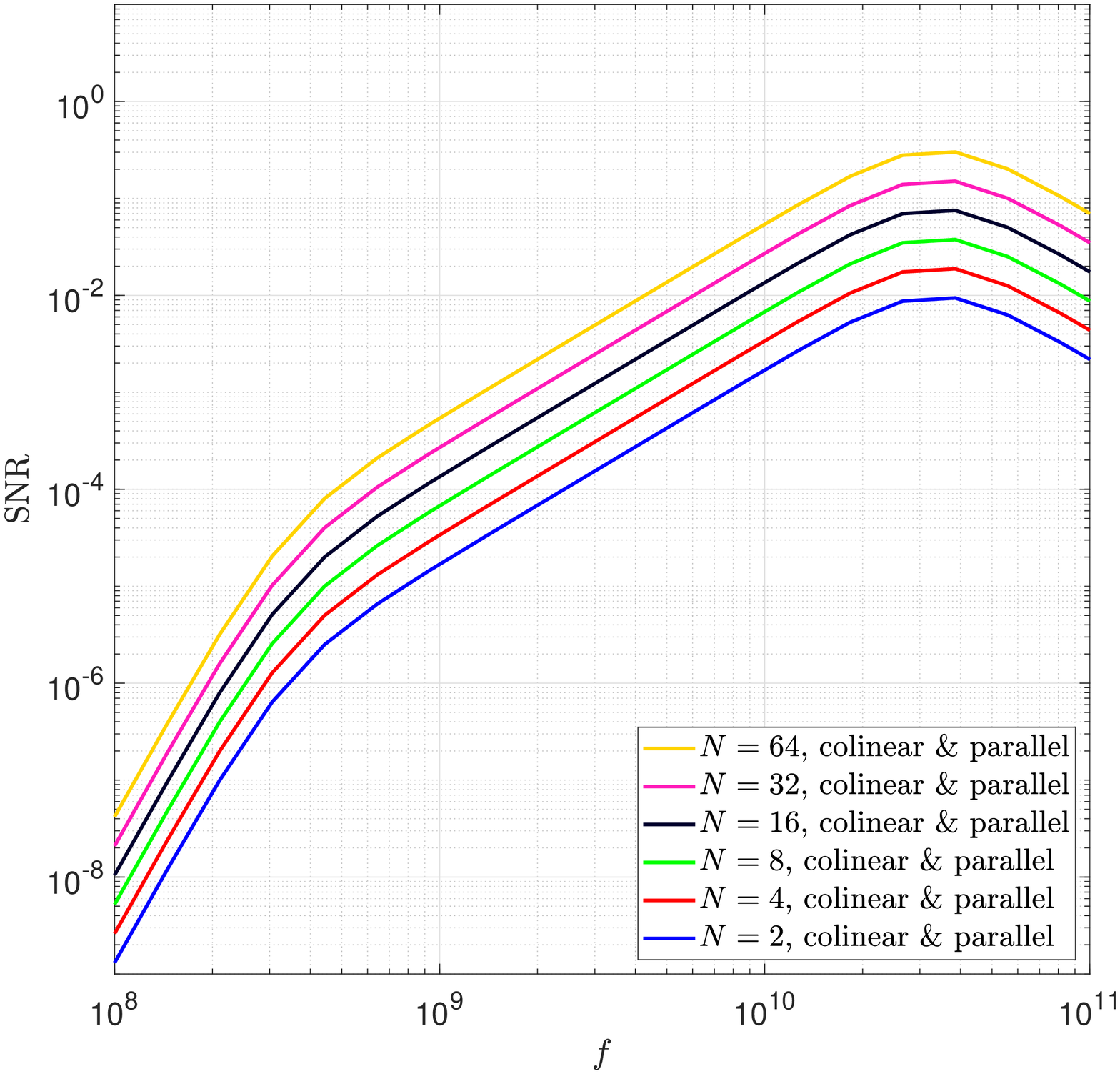}
    }}
    \caption{Plots of the MISO/SIMO SNR with optimum beamforming as a function of the frequency $f$ with $a_\textrm{T}=100\,\delta$ for SIMO, $a_\textrm{R}=100\,\delta$ for MISO.
    For both arrays, antennas are aligned either in a colinear or parallel configuration with the same fixed spacing $\delta_{\textrm{T}} = \delta_{\textrm{R}} = \delta = 0.5 [\textrm{cm}]$, and $P_{\textrm{max}} = 2\, [\text{W}]$ when the mutual coupling is (c)-(d) ignored, and (a)-(b) taken into account.}
    \label{fig:SNR-MISO-SIMO}
\end{figure*}

Fig.~\ref{fig:SNR-MISO-SIMO} depicts the SNR as a function of the frequency at the receiver side for both MISO/SIMO cases with optimum beamforming. There, to focus on the transmit side (resp. receive side) for MISO (resp. SIMO), we set the transmit (resp. receive) antenna size to $a_{\textrm{T}} = 100 \,\delta$ (resp. $a_{\textrm{R}} = 100 \,\delta$) to exclude bandwidth limitation due to the size of the transmit (resp. receive) antenna. We plot multiple SNR curves by varying the number $N$ (resp. $M$) of transmit (resp. receive) antennas for the MISO (resp. SIMO) case under both colinear and parallel configurations. When the antenna elements are weakly coupled, we observe in Figs.~\ref{fig:MISO-no-coupling} and \ref{fig:SIMO-no-coupling} an increase in the SNR values as a function of the number of transmit (resp. receive) antenna elements for both colinear and parallel configurations. This is due to antenna array gains, which is a basic result in the communication literature \cite{heath2018foundations}. While the wireless MISO/SIMO channels are reciprocal, we note a tiny SNR loss for the SIMO case in Figs. \ref{fig:SIMO-coupling} and \ref{fig:SIMO-no-coupling} as compared to the MISO case in Figs. \ref{fig:MISO-coupling} and \ref{fig:MISO-no-coupling}, respectively, due to the non-reciprocal LNAs whose impedances are not matched.

\noindent When the antenna elements are coupled as shown in Fig.~\ref{fig:MISO-coupling}, we observe in addition to the transmit/receive array gains a larger operational bandwidth for the colinear configuration as compared to the parallel configuration. This is due to the stronger radial component  of the electric ﬁeld (and hence the MC) in colinear conﬁgurations as compared to parallel conﬁgurations regardless of whether they use matching networks or not\footnote{Parallel configurations have been recently studied both with and without matching networks in \cite{akrout2021achievable} and \cite{de2020matching}, respectively.}. It is also found that the operational bandwidth of the colinear configuration becomes wider as the number of antenna elements increases (or equivalently as the MC becomes higher) leading to an almost constant current over the array aperture. This makes the colinear configuration more appealing from a geometric design perspective. Such bandwidth widening is key to enabling the broadband operation of future communication systems over multiple frequency bands. In essence, such antenna coupling-induced bandwidth gain manifests itself as a missing gain of massive MIMO which has so far been praised for antenna, diversity, and multiplexing gains only.

\begin{figure}[th!]
    \centering
    \hspace*{-0.5cm}
    \includegraphics[scale=0.37]{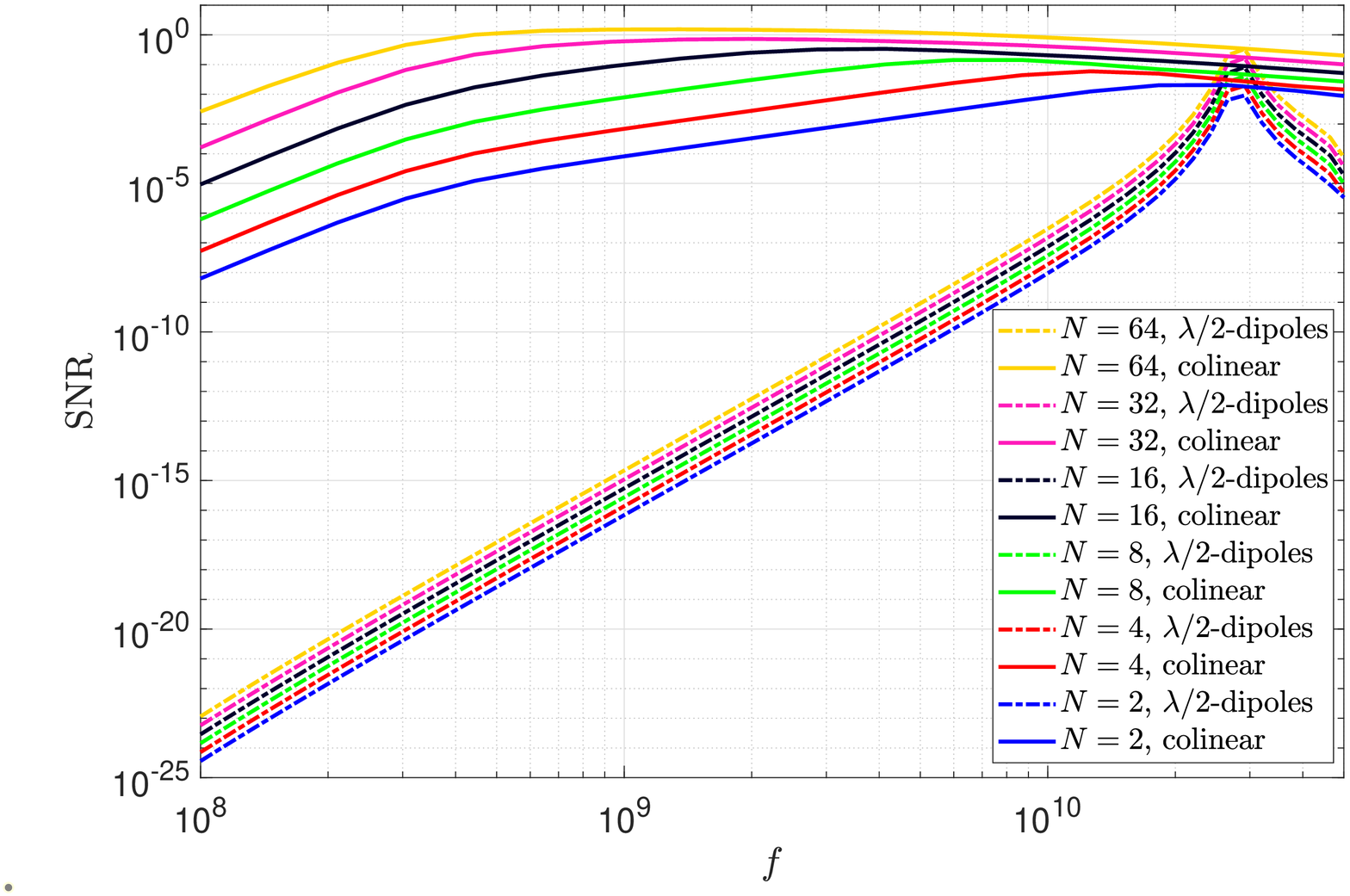}
    \caption{Plot of the MISO SNR with optimum beamforming as a function of the frequency $f$ with $a_{\textrm{R}}=100\,\delta$. Transmit antennas are either Chu's CMS antennas aligned in a colinear configuration (solid lignes) with the same fixed spacing $\delta_{\textrm{T}} = \delta_{\textrm{R}} = \delta = 0.5 [\textrm{cm}]$ or standard $\lambda/2$ dipole elements, and $P_{\textrm{max}} = 2\, [\textrm{W}]$ when the mutual coupling is taken into account.}
    \label{fig:half-lambda-dipoles}
    \vspace{-0.1cm}
\end{figure}

\noindent For more comprehensive benchmarking, Fig. \ref{fig:half-lambda-dipoles} depicts SNR plots for the linear $\lambda/2$ dipole arrays as a practical example of the standard MIMO case. We simulated dipole array impedances in Matlab using the Antenna toolbox package.

\noindent It is seen that $\lambda/2$ arrays (as in conventional MIMO) do not exhibit any bandwidth gain (i.e.,
widening) as they resonate around the center frequency (as done by each dipole separately). This is also due to their conventional parallel orientation which does not enable strong mutual coupling effects between neighboring elements as already shown in Figs.~\ref{fig:MISO-coupling} and \ref{fig:SIMO-coupling}.

\begin{figure}[h!]
    \centering
    \subcaptionbox{SIMO.\label{fig:MISO-capacity-spacing}}{{\hspace{-0.5cm}
    \includegraphics[scale=0.28]{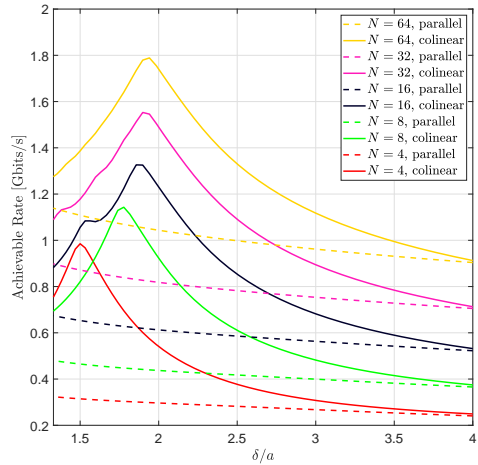}
    }}\hfill\hfill\hfill
    \subcaptionbox{MISO.\label{fig:SIMO-capacity-spacing}}{{
    \includegraphics[scale=0.28]{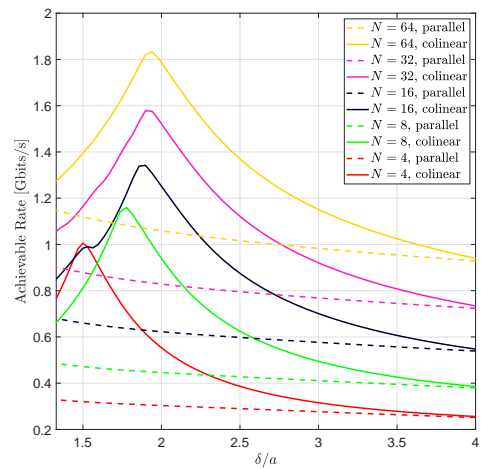}
    }}
    \caption{Plots of the SIMO/MISO achievable rate with optimum beamforming as a function of the spacing-to-antenna-size $\delta/a \in [\frac{4}{3}, 4]$ at a fixed spacing $\delta = 0.5 [\textrm{cm}]$ and $P_{\textrm{max}} = 2\, [\text{W}]$. For both arrays, antennas are aligned either in a colinear or parallel configuration for (a) SIMO, and (b) MISO communications.}
    \label{fig:MISO-MISO-capacity-spacing}
    \vspace{-0.4cm}
\end{figure}

 To corroborate the inherent symmetry of SIMO and MISO configurations, we optimize the transmit (resp. receive) antennas size $a_\textrm{T}$ (resp. $a_\textrm{R}$) for MISO (resp. SIMO). To control the MC effects, we vary the radius of the Chu spheres from the overlapping case in Fig. \ref{fig:overlapping-Chu-spheres} to the non-overlapping case in Fig. \ref{fig:non-intersecting-Chu-spheres} as a function of the fixed spacing $\delta$. More specifically, we vary the spacing-to-antenna-size ratio $\delta/a$ between $\frac{4}{3}$ and 4 as shown in Fig. \ref{fig:MISO-MISO-capacity-spacing}. There, we show the SIMO/MISO achievable rates as a function of $\delta/a$ for different numbers of transmit/receive antenna elements, namely $(N,M)\in\{4,8,16,32,64\}^2$, for both colinear and parallel configurations.
 
\noindent It is seen that the MC in the parallel configuration does not impact much the achievable rate because this configuration does not exhibit a bandwidth gain as recognized from Fig. \ref{fig:SNR-MISO-SIMO}. However, for the colinear configuration, we not only observe an increase in the achievable rate as a function of the number of antenna elements but also the presence of an optimum spacing-to-antenna-size ratio $(\delta/a)^{\tiny{\starletfill}}<2$ that maximizes the achievable rate. This ratio corresponds to the case of neighboring antenna elements with an optimum overlap between their Chu's spheres as depicted in Fig. \ref{fig:overlapping-Chu-spheres}, or equivalently an optimum positive impact of the MC effects on the achievable rate. As the overlap increases \big(i.e., backing off from the asymptotically optimum ratio  $(\delta/a)^{\tiny{\starletfill}}$, e.g., $\delta/a \in \left[4/3,(\delta/a)^{\tiny{\starletfill}}\right[$\big), the achievable rate decreases and the MC becomes harmful. Moreover, decreasing the overlap to non-intersecting Chu's spheres \big(i.e., $\delta/a \in \left[(\delta/a)^{\tiny{\starletfill}},4\right[$\big) progressively attenuates the MC effects and ultimately leads to weakly coupled neighboring antenna elements. For this reason, it is observed that parallel and colinear configurations have similar achievable rates around $\delta/a \approx 4$ at which the MC effects are almost absent. The latter scenario is similar to the widely adopted half-wavelength inter-antenna spacing (i.e., $\delta/\lambda=1/2$) for FF communications. Furthermore, we notice that the optimum spacing-to-antenna-size ratio increases as the number of transmit antennas increases and ultimately reaches the asymptotic value $1.9 \approx \sqrt[3]{6\,\zeta(3)}$, thereby corroborating the condition for tight coupling we established in Result~\ref{result:condition-tight-coupling}. It also indicates that MC becomes less critical as the number of antennas increases since it can still lead to nearly constant currents across frequencies between neighboring antenna elements when the antenna array is large enough. This is in agreement with a previous observation that stipulates the existence of an optimum spacing that maximizes the array gain under tightly-coupled antenna elements \cite{ivrlavc2010high}. \vspace{-0.2cm}
\subsection{MIMO achievable rate}
\begin{figure*}[h!]
    \centering
    \subfloat[\hspace{0cm}colinear/parallel \textit{weakly} coupled antennas.\label{fig:MIMO-LOS-no-precoding-no-coupling-colinear-parallel}]{{
    \includegraphics[scale=0.375]{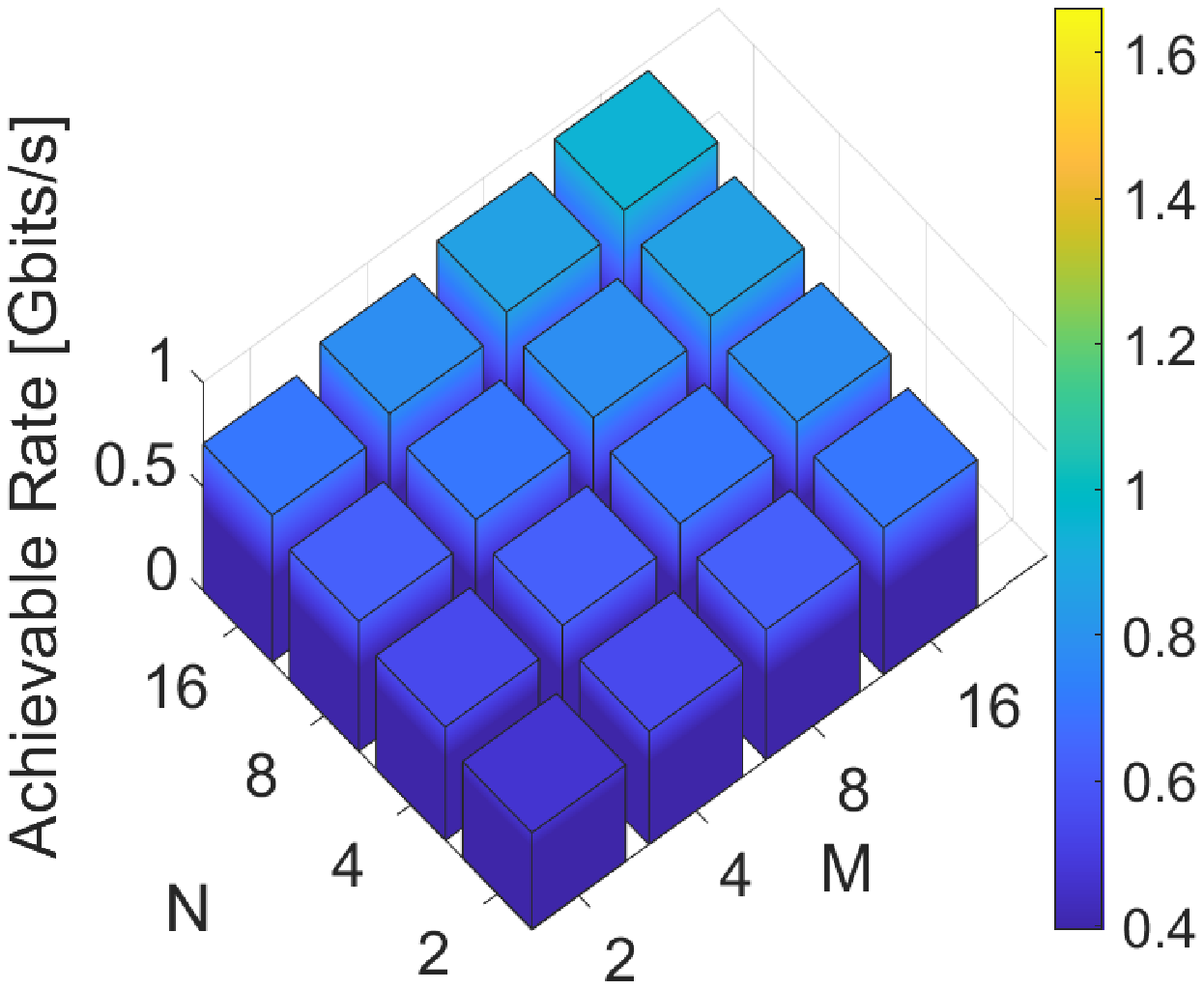}
    }}
    \subfloat[colinear \textit{tightly} coupled antennas.\label{fig:MIMO-LOS-no-precoding-coupling-colinear}]{{
    \includegraphics[scale=0.375]{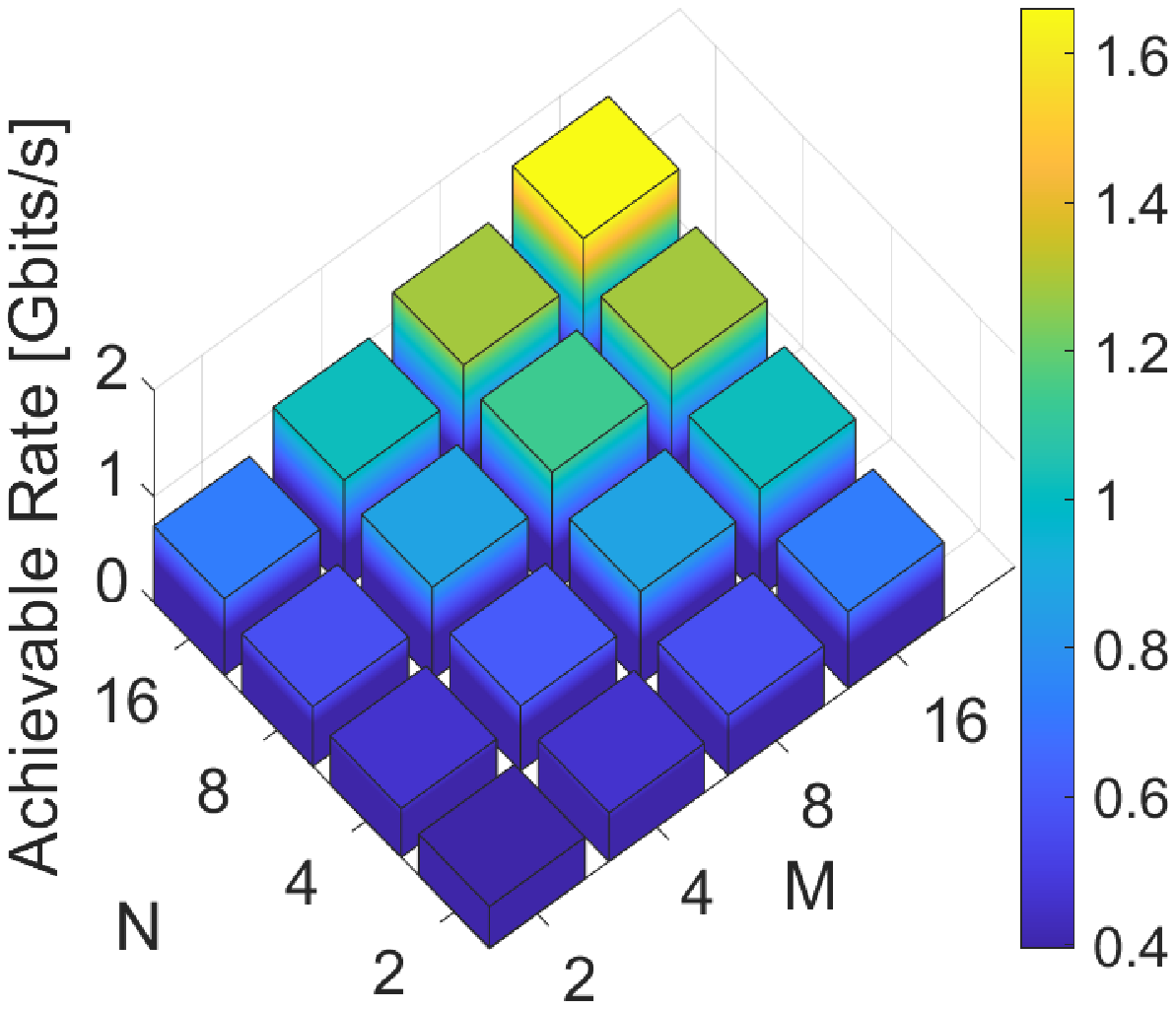}
    }}
    \subfloat[parallel \textit{tightly} coupled antennas.\label{fig:MIMO-LOS-no-precoding-coupling-parallel}]{{
    \includegraphics[scale=0.375]{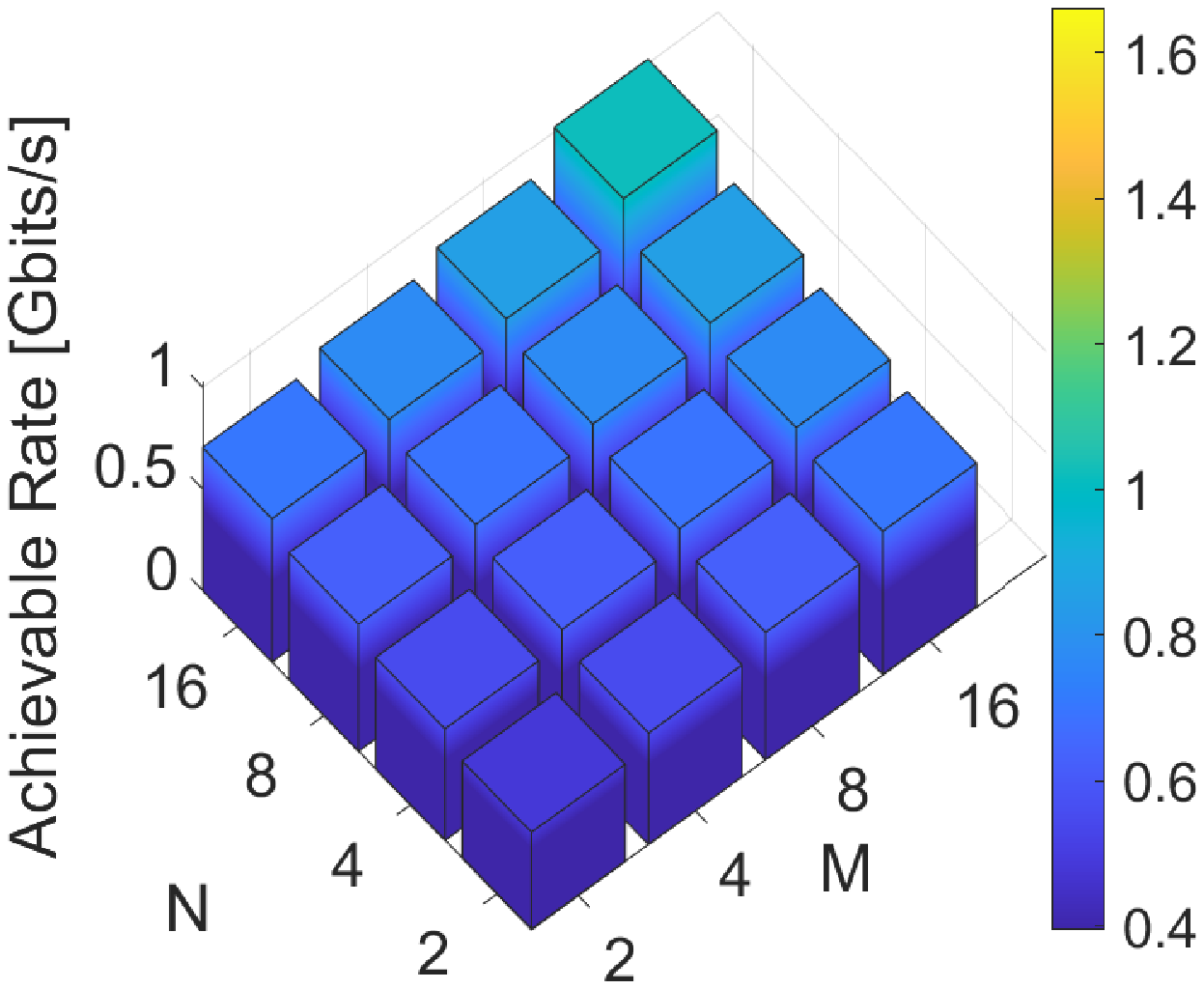}
    }}
    \caption{Heatmaps of the achievable rate over LoS channels as a function of the number of transmit and receive antenna elements $N$ and $M$. For both arrays, antennas are aligned either in a colinear or parallel configuration with the same fixed spacing $\delta_{\textrm{T}} = \delta_{\textrm{R}} = \delta = 0.5 [\textrm{cm}]$, $a_\textrm{T}=a_\textrm{R}=\delta/\sqrt[3]{6 \,\zeta(3)}$, and $P_{\textrm{max}} = 2\, [\text{W}]$ when the mutual coupling is (a) ignored, and (b)-(c) taken into account.}
    \label{fig:MIMO-capacity-LOS} 
    \vspace{-0.3cm}
\end{figure*}

\begin{figure*}[h!]
    \centering
    \vspace{-0.2cm}
    \subfloat[colinear/parallel \textit{weakly} coupled antennas.\label{fig:MIMO-Rayleigh-no-precoding-no-coupling-colinear-parallel}]{{
    \includegraphics[scale=0.375]{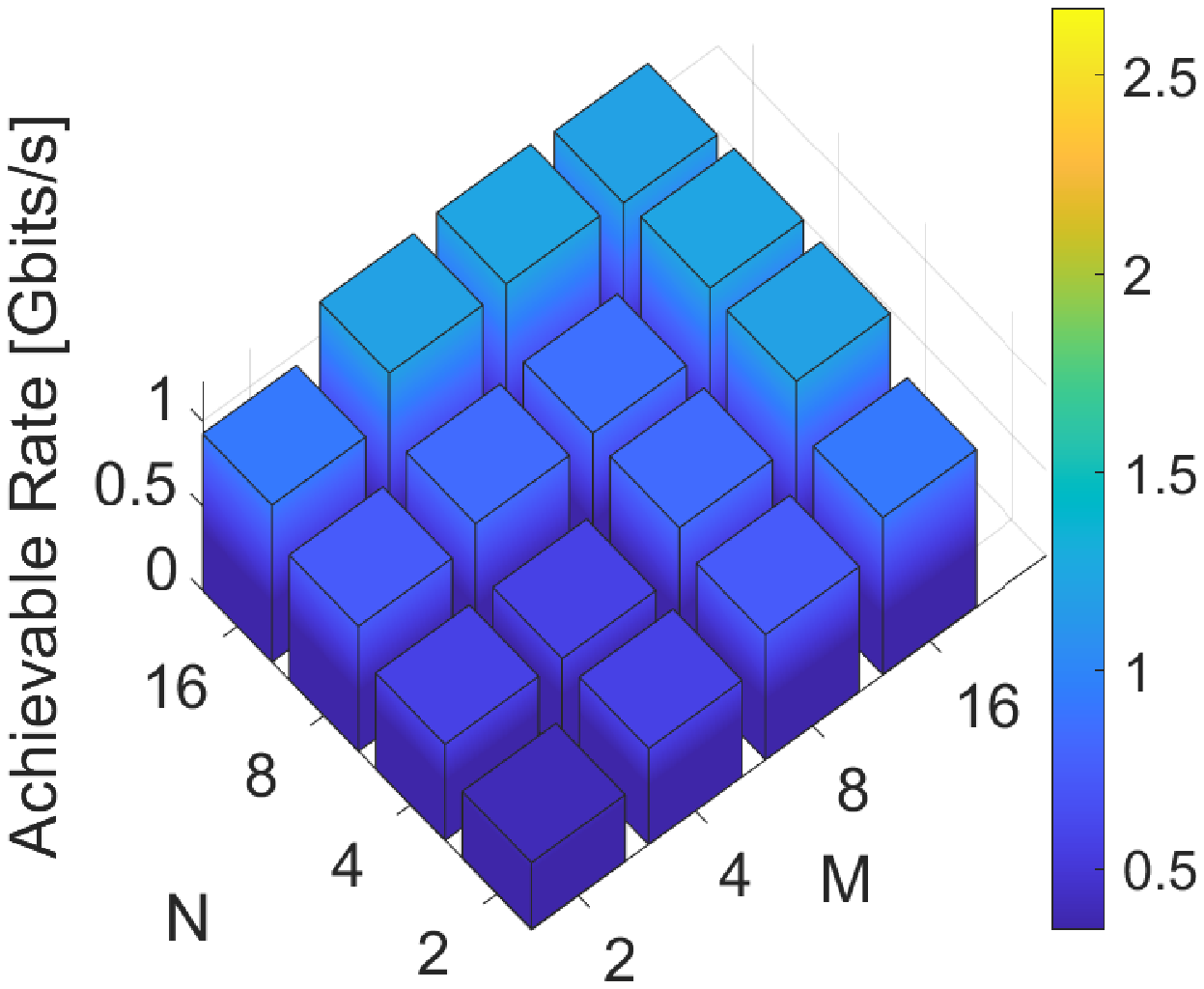}    }}
    \subfloat[colinear \textit{tightly} coupled antennas.\label{fig:MIMO-Rayleigh-no-precoding-coupling-colinear}]{{
     \includegraphics[scale=0.375]{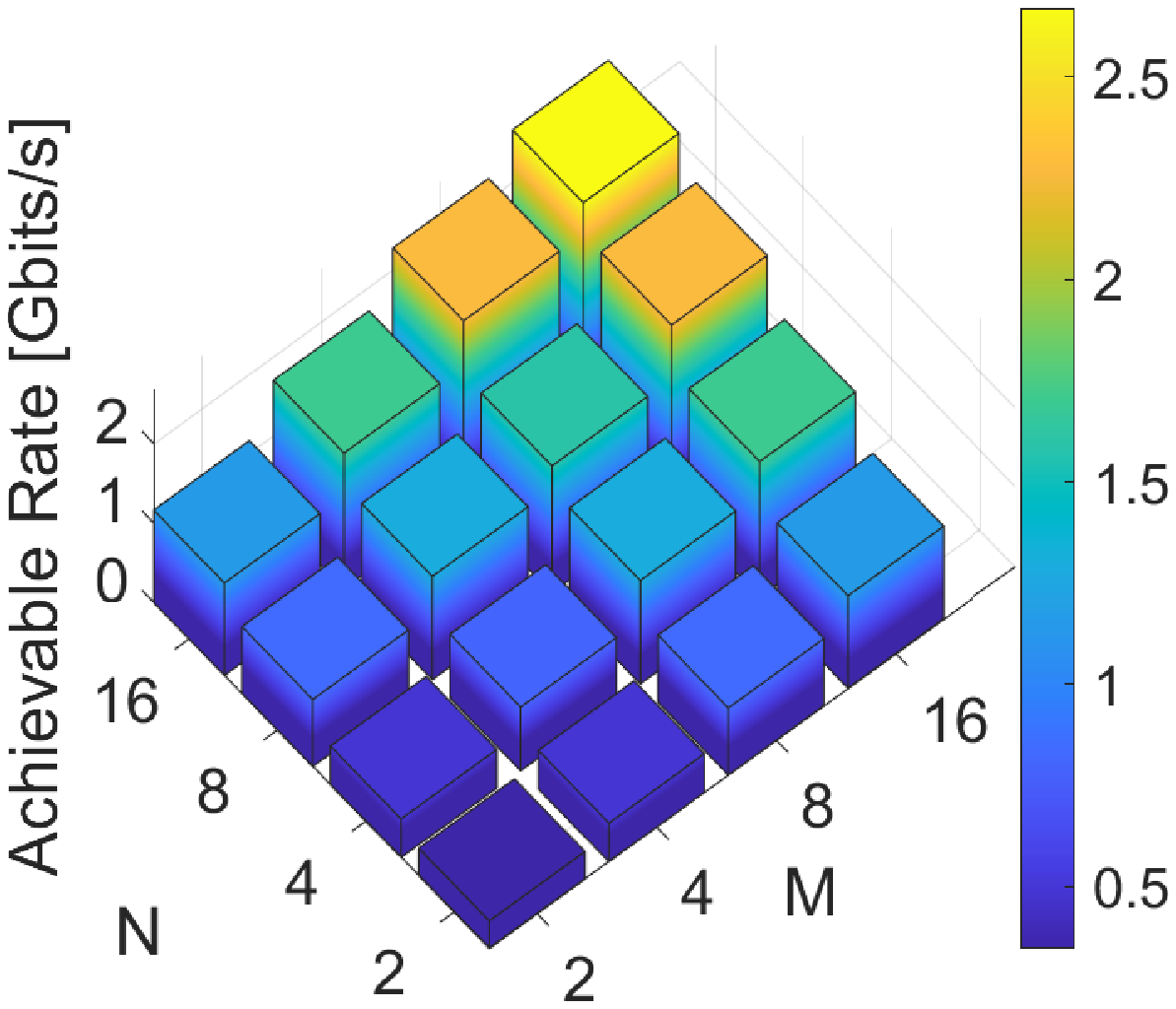}
    }}
    \subfloat[parallel \textit{tightly} coupled antennas.\label{fig:MIMO-Rayleigh-no-precoding-coupling-parallel}]{{    
    \includegraphics[scale=0.375]{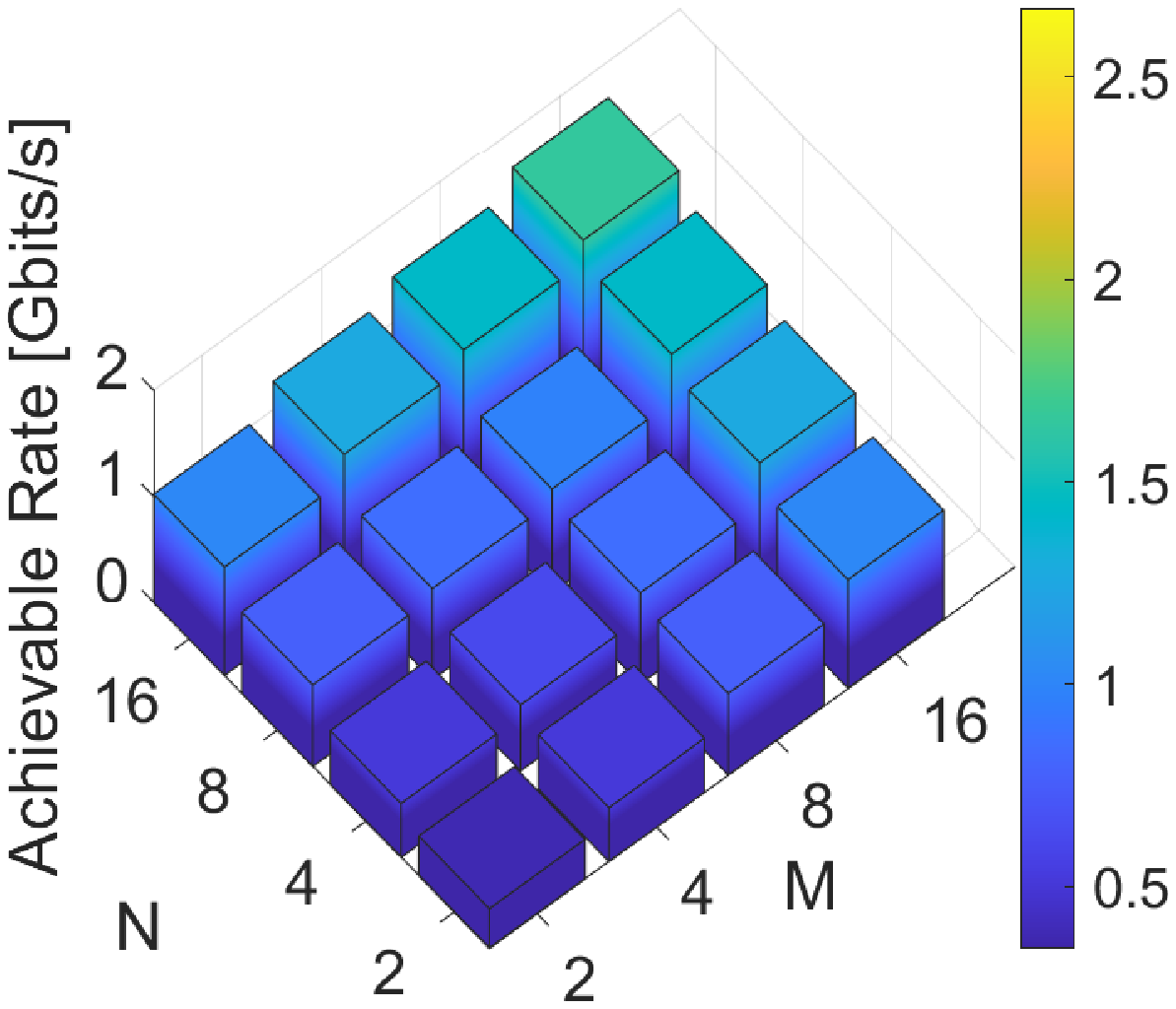}
    }}
    
    \caption{Heatmaps of the achievable rate over Rayleigh channels as a function of the number of transmit and receive antenna elements $N$ and $M$. For both arrays, antennas are aligned either in a colinear or parallel configuration with the same fixed spacing $\delta_{\textrm{T}} = \delta_{\textrm{R}} = \delta = 0.5 [\textrm{cm}]$, $a_\textrm{T}=a_\textrm{R}=\delta/\sqrt[3]{6 \,\zeta(3)}$, and $P_{\textrm{max}} = 2\, [\text{W}]$ when the mutual coupling is (a) ignored, and (b)-(c) taken into account.}
    \label{fig:MIMO-capacity-Rayleigh}
\end{figure*}

Figs. \ref{fig:MIMO-capacity-LOS} and \ref{fig:MIMO-capacity-Rayleigh} show the heatmaps of the MIMO achievable rate over LoS and Rayleigh channels, respectively, as a function of the number of transmit and receive antennas $N$ and $M$ for both colinear and parallel configurations.

\noindent In the case of weakly coupled antenna elements, we observe in Figs.~\ref{fig:MIMO-LOS-no-precoding-no-coupling-colinear-parallel} and \ref{fig:MIMO-Rayleigh-no-precoding-no-coupling-colinear-parallel} that the achievable rate improves very slowly with the number of transmit and receive antennas $N$ and $M$. This result is common knowledge for LoS channels and also holds for low-SNR Rayleigh channels, which is usually the case in FF communications.
\begin{figure}[h!]
    \centering
    \vspace{-0.2cm}
    \subfloat[colinear antennas.\label{fig:MIMO-LOS-no-precoding-coupling-colinear-high-SNR}]{{
    \includegraphics[scale=0.3]{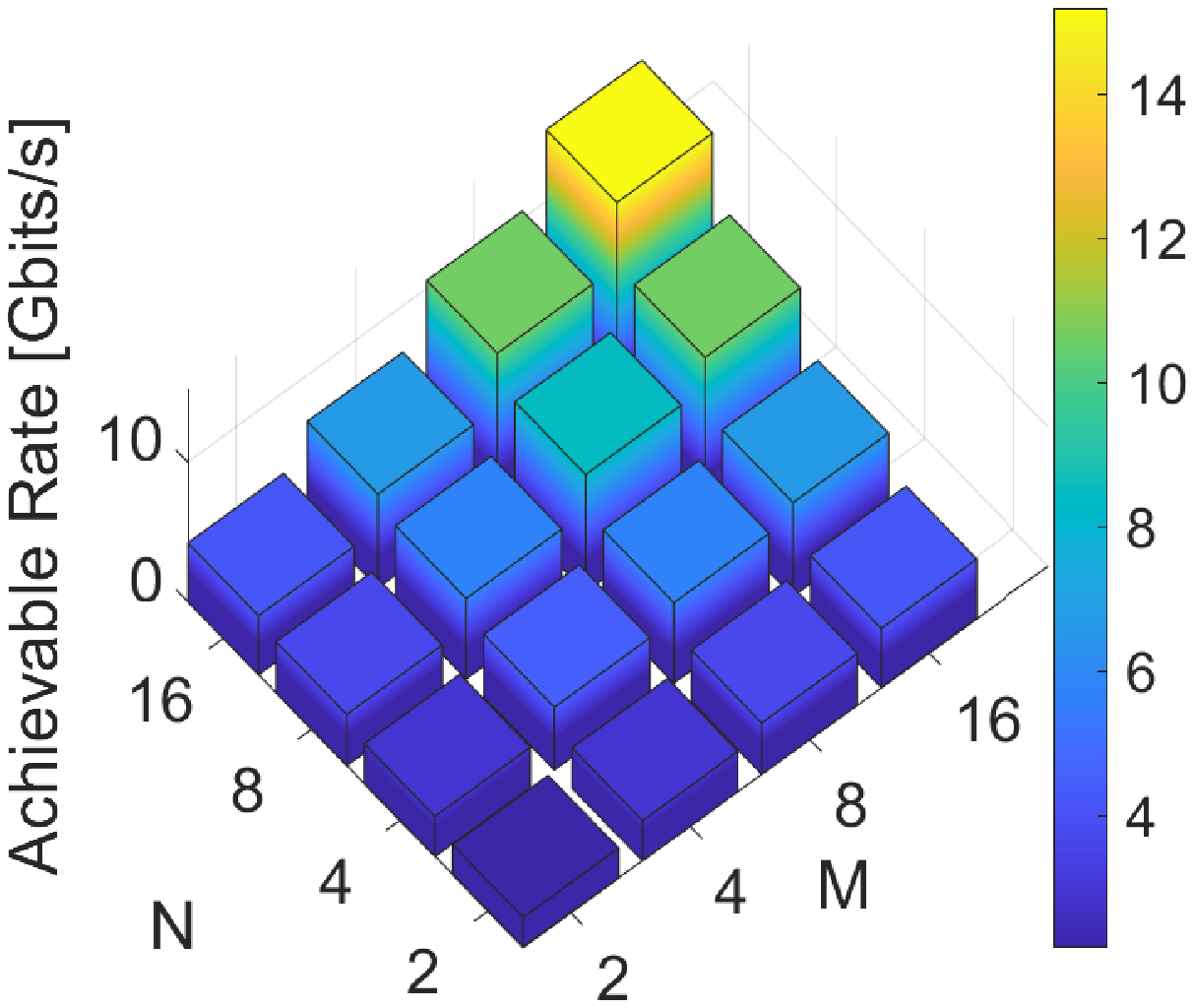} 
    }}
    \subfloat[parallel antennas.\label{fig:MIMO-LOS-no-precoding-coupling-parallel-high-SNR}]{{ 
    \includegraphics[scale=0.3]{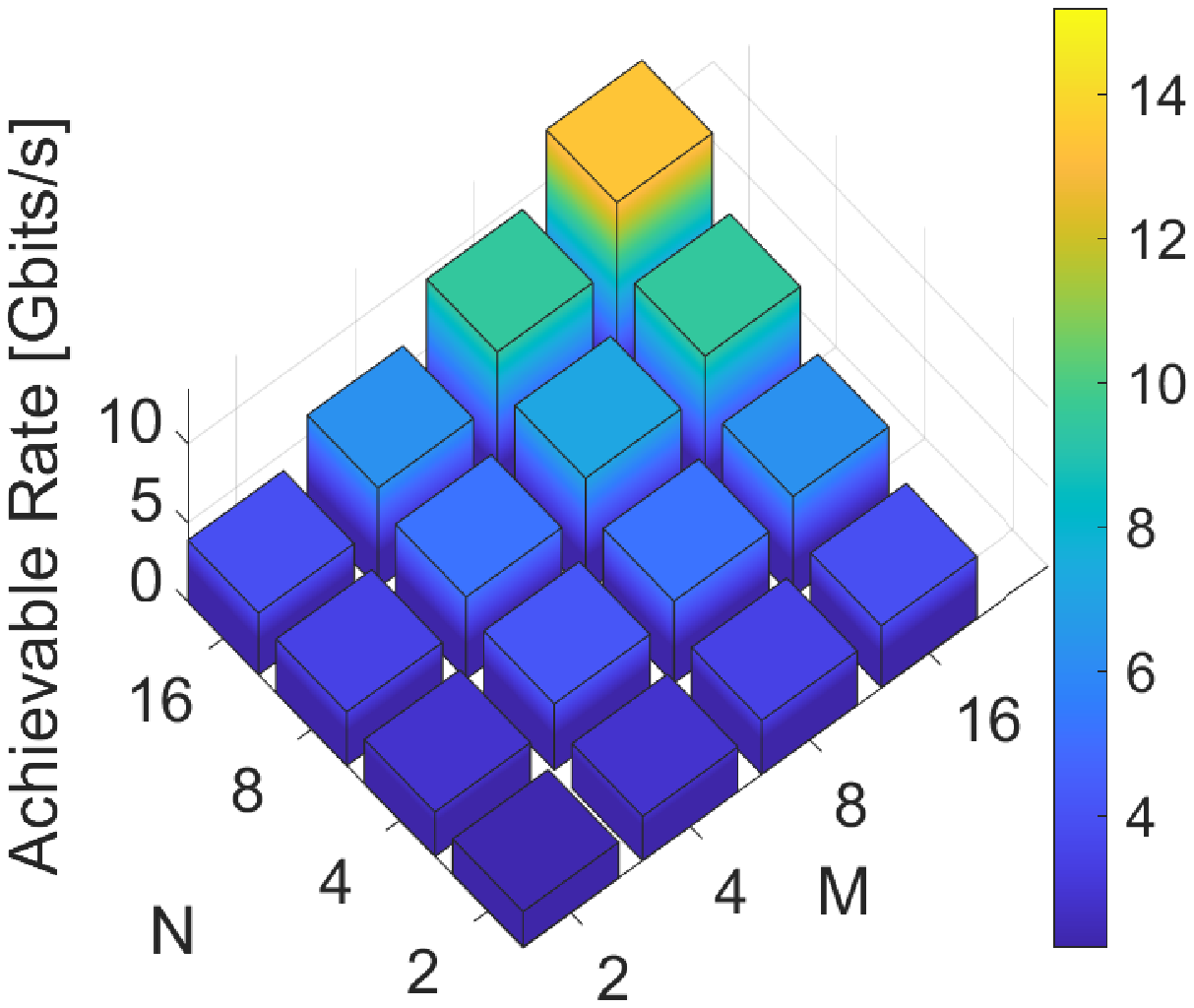}
    }}
    \caption{Heatmaps of the achievable rate over Rayleigh channels as a function of the number of transmit and receive antenna elements $N$ and $M$. For both arrays, antennas are aligned either in a (a) colinear or (b) parallel configuration with the same fixed spacing $\delta_{\textrm{T}} = \delta_{\textrm{R}} = \delta = 0.5 [\textrm{cm}]$, $a_\textrm{T}=a_\textrm{R}=\delta/\sqrt[3]{6 \,\zeta(3)}$, and $P_{\textrm{max}} = 4000\, [\text{W}]$ when the mutual coupling is taken into account.}
    \label{fig:MIMO-capacity-LOS-high-SNR}
\end{figure}

\noindent In the high-SNR regime, however, the achievable rate over Rayleigh channels depends on the channel rank $R=\textrm{min}(M,N)$ \cite{heath2018foundations}. To confirm this fact, we increase the total power, $P_{\textrm{max}}$, by an unrealistic factor of 2000 to operate in the high-SNR regime. The latter is not practical for FF communications but is still useful to validate the simulated achievable rate with the theoretical one. We report the obtained results in Fig. \ref{fig:MIMO-capacity-LOS-high-SNR} for both the colinear and parallel antenna configurations. There, the achievable rate approximately exhibits the theoretical dependence on the rank $R=\textrm{min}(M,N)$. We also note that the achievable rates are almost the same for both configurations because the SNR gain vanishes at the high-SNR regime and so does the impact of the geometric antenna alignment.

\noindent By further comparing Fig. \ref{fig:MIMO-LOS-no-precoding-coupling-colinear} and Fig. \ref{fig:MIMO-LOS-no-precoding-coupling-parallel} for LoS channels and Fig. \ref{fig:MIMO-Rayleigh-no-precoding-coupling-colinear} and Fig. \ref{fig:MIMO-Rayleigh-no-precoding-coupling-parallel} for Rayleigh channels, it becomes clear that MC improves the achievable rate for the colinear configuration only. This stems from the bandwidth gain already observed for the colinear configuration only in Fig.~\ref{fig:SNR-MISO-SIMO}.

Furthermore, by comparing Figs. \ref{fig:MIMO-LOS-no-precoding-coupling-colinear} and \ref{fig:MIMO-LOS-no-precoding-coupling-parallel} \big(resp. Figs. \ref{fig:MIMO-Rayleigh-no-precoding-coupling-colinear} and \ref{fig:MIMO-Rayleigh-no-precoding-coupling-parallel}\big) with Fig. \ref{fig:MIMO-LOS-no-precoding-no-coupling-colinear-parallel} \big(resp. Fig. \ref{fig:MIMO-Rayleigh-no-precoding-no-coupling-colinear-parallel}\big), it is seen again that MC significantly increases the achievable rate of couple massive MIMO systems. This result is opposed to the negative percept of MC in the open literature which advocates that MC substantially degrades the performance \cite{chen2018review,janaswamy2002effect}. The reason for this disparity lies in the coupling models used in prior art which relies on isotropic coupling expressions based on Bessel functions\cite{bird2021mutual}. These coupling models are simplistic as they account for the MC effects under the idealized assumption of isotropic radiators while disregarding the coupling caused by noise correlation, unlike our mutual coupling model $\mathcal{MC}(d, \beta, \gamma)$ in (\ref{eq:mutual-coupling-SISO}).

\begin{figure}[h!]
    \centering
    \includegraphics[scale=0.36]{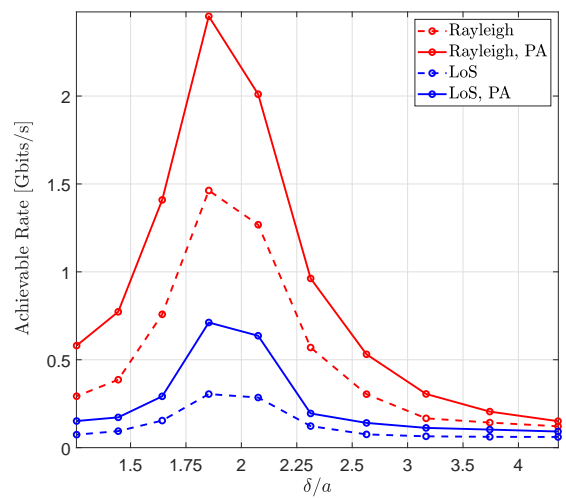}
    \caption{Achievable rate over both LoS (in blue) and Rayleigh channels (in red) under uniform and optimum power allocation (PA) strategies as a function of the spacing-to-antenna-size $\delta/a$ at $P_{\textrm{max}} = 2\, [\text{W}]$. Antenna elements within transmit and receive arrays are aligned in a colinear configuration with the same number of antennas $N=M=16$, same spacing $\delta_{\textrm{T}} = \delta_{\textrm{R}} = \delta = 0.5 [\textrm{cm}]$, and same antenna size $a_\textrm{T}=a_\textrm{R}=\delta/\sqrt[3]{6 \,\zeta(3)}$.}
    \label{fig:MIMO-capacity-spacing}
\end{figure}

\noindent Finally, we examine in Fig.~\ref{fig:MIMO-capacity-spacing} the impact of MC on the MIMO achievable rate with the colinear configuration using $M=N=16$ antennas under $i)$ uniform and optimum precoding strategies and $ii)$ LoS and Rayleigh channels. There, it is observed that with optimal precoding the achievable rate under Rayleigh channels is significantly higher than under LoS channels. It is also seen how the gain provided by the optimum PA strategy is non-uniform but is rather highly dependent on the spacing-to-antenna-size ratio and hence on the MC between neighboring antenna elements. This result demonstrates how the physically consistent modeling approach bridges the design gaps between antenna and communication considerations, e.g., antenna spacing and channel type, respectively.

\clearpage
\section{Conclusion}
In this paper, we characterized the achievable rate of tightly-coupled massive MIMO wireless systems for FF communication under physical limitations on the antenna size while taking the mutual coupling within the transmit/receive arrays into account. By employing a circuit-theoretic approach, we incorporated the effects of mutual coupling and the restrictions on Chu's CMS antennas in the MIMO channel response and the noise correlation. We then derived the impedances of the transmit/receive arrays whose antenna elements are aligned either in a colinear or a parallel configuration, as well as, the far-field transimpedances for both line-of-sight and Rayleigh channels. This led to a full characterization of the input/output relationship of the proposed circuit-equivalent MIMO model. Unlike the simplistic assumptions in prior art, we found that our physically-consistent model reveals beneficial MC effects. In particular we showed that appropriately accounting for MC leads to a bandwidth gain for the colinear antenna array configuration only, thereby providing an information-theoretic justification to the broadband behavior of existing connected array designs. We also established the condition for tight coupling in uniform linear Chu's CMS arrays in terms of the optimum spacing-to-antenna-size ratio (and hence the optimum mutual coupling level). The analysis of tightly-coupled massive MIMO presented in this paper revealed the importance of the mutual coupling effects when appropriately taken into account through a physically consistent model. Many avenues for further extension of this work are noteworthy. It is possible to consider Chu antennas radiating other modes beyond the TM mode (e.g., electromagnetic antennas). It is also possible to incorporate impedance matching networks and/or the electromagnetic properties of various propagation environments.

\begin{appendices}

\renewcommand{\thesectiondis}[2]{\Roman{section}:}
\section{Input/output relationship of a MIMO communication system}\label{appendix:input-output-mimo-channel}
To find the input-output relationship between the input voltage vector $\bm{v}_{\textrm{G}}$ and the output voltage $\bm{v}_{\textrm{L}}$, we first apply the Kirchhoff's voltage law (KVL) between the two closed loops on the left- and right-hand sides of the joint impedance matrix $\bm{Z}_{\textrm{MIMO}}$ in Fig.~\ref{fig:MIMO-communication-circuit}, thereby yielding
\begin{subequations}\label{eq:KVL-Zmimo}
    \begin{align}
    \bm{v}_{\textrm{R}} &= - \bm{\widetilde{v}}_{\textrm{N,R}} - R_{\textrm{in}}\,\bm{i}_{\textrm{R}}, \label{eq:KVL-Vr}\\
    \bm{v}_{\textrm{T}} &= \bm{v}_{\textrm{G}} - R\,\bm{i}_{\textrm{T}} - \bm{\widetilde{v}}_{\textrm{N,T}}.\label{eq:KVL-Vt}
    \end{align}
\end{subequations}
After equating the expressions of $\bm{v}_{\textrm{R}}$ in (\ref{eq:mimo-V=ZI}) and (\ref{eq:KVL-Vr}), we obtain:
\begin{equation}\label{eq:Ir-formula}
    \bm{i}_{\textrm{R}} = - (\bm{Z}_{\mathrm{R}} + R_{\textrm{in}}\,\mathbf{I}_{M\times M})^{-1}\,(\bm{\widetilde{v}}_{\textrm{N,R}}  + \bm{Z}_{\mathrm{RT}} \,\bm{i}_{\textrm{T}}).
\end{equation}
By recalling (\ref{eq:LNA-voltage-gain}) and using the fact that $\bm{i}_{R_{\textrm{in}}}(f) = - \bm{i}_{\textrm{R}}(f)$, we also apply KVL in the LNA loop to get:
\begin{equation}\label{eq:output-voltage-formula1}
    \bm{v}_{\textrm{L}} = \bm{v}_{\textrm{N}} + \beta\,R_{\textrm{in}}\,(\bm{Z}_{\mathrm{R}} + R_{\textrm{in}}\,\mathbf{I}_{M\times M})^{-1}\,(\bm{\widetilde{v}}_{\textrm{N,R}}  + \bm{Z}_{\mathrm{RT}} \,\bm{i}_{\textrm{T}}).
\end{equation}
Using (\ref{eq:Ir-formula}) and the expression of $\bm{v}_{\textrm{T}}$ in (\ref{eq:mimo-V=ZI}), it follows that:
\begin{equation}\label{eq:Vt-formula}
    \bm{v}_{\textrm{T}} = \bm{Z}_{\mathrm{T}}\,\bm{i}_{\textrm{T}} - \bm{Z}_{\mathrm{TR}} \, (\bm{Z}_{\mathrm{R}} + R_{\textrm{in}}\,\mathbf{I}_{M\times M})^{-1}\,(\bm{\widetilde{v}}_{\textrm{N,R}}  + \bm{Z}_{\mathrm{RT}} \,\bm{i}_{\textrm{T}}).
\end{equation}
Now, equating (\ref{eq:Vt-formula}) and (\ref{eq:KVL-Vt}) gives
\begin{equation}
\begin{aligned}
    \bm{i}_{\textrm{T}} &= \left(\bm{Z}_{\mathrm{T}} + R\,\mathbf{I}_{N\times N}- \bm{Z}_{\mathrm{TR}}\,(\bm{Z}_{\mathrm{R}} + R_{\textrm{in}}\,\mathbf{I}_{M\times M})^{-1}\,\bm{Z}_{\mathrm{RT}}\right)^{-1}\\
    &\hspace{1.3cm}\times\left(\bm{v}_{\textrm{G}} - \bm{\widetilde{v}}_{\textrm{N,T}} + \bm{Z}_{\mathrm{TR}} \,(\bm{Z}_{\mathrm{R}} + R_{\textrm{in}}\,\mathbf{I}_{M\times M})^{-1}\,\bm{\widetilde{v}}_{\textrm{N,R}}\right),
\end{aligned}
\end{equation}
which, when injected back in (\ref{eq:output-voltage-formula1}), yields the desired relationship between the input voltage $\bm{v}_{\textrm{G}}$ and output voltage $\bm{v}_{\textrm{L}}$:
\begin{equation}
    \begin{aligned}[b]
    \bm{v}_{\textrm{L}} &= \bm{v}_{\textrm{N}} + \beta\,R_{\textrm{in}}\,\bm{P}\left(\bm{\widetilde{v}}_{\textrm{N,R}}+\bm{Z}_{\mathrm{RT}}\,\bm{Q}\left(\bm{v}_{\textrm{G}}-\bm{\widetilde{v}}_{\textrm{N,T}} + \bm{Z}_{\mathrm{TR}}\,\bm{P}\, \bm{\widetilde{v}}_{\textrm{N,R}}\right)\right),
    \end{aligned}
\end{equation}
where $\bm{P}$ and $\bm{Q}$ are given in (\ref{eq:P-and-Q}).

\section{Derivation of the far-field Line-of-Sight transimpedance}\label{appendix:FF-mutual-impedance}
We resort to basic circuit theory analysis of the FF SISO communication model described in Fig.~\ref{fig:siso-far-field-system-model} to find the FF mutual impedance, $Z_{\mathrm{RT},mn}(f)$. Using the current divider, $i_{R_1,n}(f)$ and $i_{\textrm{T},n}(f)$ are related as follows:
\begin{equation}\label{eq:relations-current-dividers}
    i_{R_1,n}(f)=i_{\textrm{T},n}(f)\,\left( \frac{\textrm{j}2\pi fa_{\textrm{T}}}{c + \textrm{j}2\pi fa_{\textrm{T}}} \right)\,[\text{A}].
\end{equation}
To find the transmit-receive impedance $Z_{{\mathrm{RT},nm}}(f)$ that relates the receiver voltage $v_{\textrm{R},m}(f)$ and the transmitter current $i_{\textrm{T},n}(f)$, we first open-circuit the receiver current by setting $i_{\textrm{R},m}(f) = 0$. The impedance $Z_{R_2\,\parallel\,L}$ of the parallel-connected resistor $R_2$ and the inductance $L = a_{\textrm{R}}\,R_2/c$ is then given by:
\begin{equation}
    Z_{R_2\,\parallel\,L}(f) = \frac{\textrm{j}2\pi fa_{\textrm{R}}\,R_2}{c + \textrm{j}2\pi f a_{\textrm{R}}}\,[\Omega].
\end{equation}
By recalling the expression of $|i_s(f)|$ in (\ref{eq:friss-current-magnitude}) and using Ohm's law for the impedance $Z_{R_2\,\parallel\,L}(f)$, i.e., $v_{\textrm{R},m}(f) = Z_{R_2\,\parallel\,L}(f) ~ i_s(f)$, one obtains:
\begin{equation}\label{eq:V2-formula}
\begin{aligned}[b]
    v_{\text{R},m}(f) &= i_{R_1,n}(f)\,\frac{c}{2{\pi}fd^{\frac{\alpha}{2}}}\,\sqrt{\frac{G_\textrm{T} G_\textrm{R} R_1}{R_2}}\\
    &\hspace{1cm}\times\left(\frac{\textrm{j}2{\pi}f a_{\textrm{R}}\, R_2}{c+\textrm{j}2\pi f a_{\textrm{R}}}\right)\,e^{-\textrm{j}\phi_{mn}}\,[\text{V}],
\end{aligned}
\end{equation}
\noindent where $\alpha$ is the path-loss exponent and $\phi_{mn} = \frac{2\pi f}{c} \big[(n-1)\,\delta_{\textrm{T}}\,\cos{\theta_{\textrm{T}}} + (m-1)\, \delta_{\textrm{R}}\,\cos{\theta_{\textrm{R}}}\big]$ is the relative phase between receive antenna $m$ and transmit antenna $n$. Here, $\theta_{\textrm{T}}$ and $\theta_{\textrm{R}}$ are the angle of arrival and departure defined w.r.t. the broadside axis of the transmit and receive arrays, respectively.

\noindent Injecting (\ref{eq:relations-current-dividers}) into (\ref{eq:V2-formula}) yields:
\begin{equation}
    \begin{aligned}[b]
    v_{\text{R},m}(f) &= i_{\text{T},n}(f)\,\frac{c}{2{\pi}fd^{\frac{\alpha}{2}}}\left( \frac{\textrm{j}2\pi f a_{\textrm{T}}\,\sqrt{R_1}}{c + \textrm{j}2\pi f a_{\textrm{T}}} \right)\,\sqrt{G_\text{T} G_\text{R}}\\
    &\hspace{1cm}\times\left(\frac{\textrm{j}2\pi f a_{\textrm{R}} \sqrt{R_2}}{c+\textrm{j}2\pi f a_{\textrm{R}}}\right)\,e^{-\textrm{j}\phi_{mn}}\,[\text{V}],
    \end{aligned}
\end{equation}
\noindent from which we obtain the expression of the $mn$th transimpedance element $Z_{\mathrm{RT},mn}(f)$ as
\begin{equation}\label{eq:FF-mutual-impedance-line-of-sight-appendix}
\begin{aligned}
    Z_{\mathrm{RT},mn}^{\textrm{LoS}}(f) &= \frac{v_{\text{R},m}(f)}{i_{\text{T},n}(f)}= \frac{c}{2{\pi}fd^{\frac{\alpha}{2}}}\,\left( \frac{\textrm{j}2\pi f a_{\textrm{T}}\,\sqrt{R_1}}{c + \textrm{j}2\pi f a_{\textrm{T}}} \right)\sqrt{G_\text{T} G_\text{R}}\\
    & \hspace{1cm} \times \left(\frac{\textrm{j}2\pi f a_{\textrm{R}} \sqrt{R_2}}{c+\textrm{j}2\pi f a_{\textrm{R}}}\right)\,e^{-\textrm{j}\phi_{mn}}\,[\Omega].
    \end{aligned}
\end{equation}
Using the expressions of the impedances $\bm{Z_{\textrm{R}}}$ and $\bm{Z_{\textrm{T}}}$ from Section \ref{subsec:self-impedances-matrices}, one can write the LoS transimpedance matrix $\bm{Z}_{\mathrm{RT}}^{\textrm{LoS}}(f)$ as follows:
\begin{equation}\label{eq:FF-mutual-impedance-line-of-sight-matrix-appendix}
    \begin{aligned}
    \bm{Z}_{\mathrm{RT}}^{\textrm{LoS}}(f) &= \frac{c\,\sqrt{G_\text{T}\,G_\text{R}}}{2{\pi}fd^{\frac{\alpha}{2}}}~\textrm{diag}\left(\Re{\{\bm{Z_{\textrm{R}}}(f)\}}\right)^{\frac{1}{2}}\,\bm{a}_\mathrm{R}(\theta_{\mathrm{R}})\,\bm{a}_{\mathrm{T}}^{\mathsf{T}}(\theta_{\mathrm{T}})\\
    &\hspace{1cm}\times\textrm{diag}\left(\Re{\{\bm{Z_{\textrm{T}}}(f)\}}\right)^{\frac{1}{2}}\,e^{-j\phi}~[\Omega],
    \end{aligned}
\end{equation}
where $\phi$, $\bm{a}_\mathrm{T}(\theta_{\mathrm{T}})$, and $\bm{a}_\mathrm{R}(\theta_{\mathrm{R}})$  are given in (\ref{eq:relative-phase}), (\ref{eq:steering-vector-Tx}) and (\ref{eq:steering-vector-Rx}), respectively.

\section{Proof of the condition for tight coupling}\label{appendix:proof-result1}
We consider a tightly-coupled massive MIMO system with transmit/receive arrays consisting of $N$ and $M$ antennas, respectively, that are located at their mutual FF region. We write Ohm's law applied to the transmit impedance matrix $\bm{Z}_{\text{T}}(f)$ as:\vspace{-0.5cm}

\begin{subequations}\label{appendix:ohm-law-elementwise}
\small
\begin{align}
    \bm{v}_{\textrm{T}}(f) &= \bm{Z}_{\text{T}}(f)\, \bm{i}_{\textrm{T}}(f)\\
            &=
\begin{bNiceArray}{cccc}[margin]
 Z_{\text{T},11}^{\textrm{self}} & \cdots  & \cdots & Z_{\text{T},1N}^{\textrm{mutual}}\\
 Z_{\text{T},21}^{\textrm{mutual}}    &  Z_{\text{T},22}^{\textrm{self}}     & \cdots & Z_{\text{T},2N}^{\textrm{mutual}}\\
 \vdots      &  \vdots    & \ddots & \vdots   \\
  Z_{\text{T},N1}^{\textrm{mutual}}   &  \cdots    & \cdots & Z_{\textrm{T},NN}^{\textrm{self}}
\end{bNiceArray} \, \bm{i}_{\textrm{T}}(f). \label{appendix:ohm-law-elementwise-b}
\end{align}
\end{subequations}

\noindent where the dependence on the frequency argument ($f$) was dropped in (\ref{appendix:ohm-law-elementwise-b}) for notational conveniences. In (\ref{appendix:ohm-law-elementwise-b}), each $k$th row in $\bm{Z}_{\text{T}}(f)$ is composed of:
\begin{itemize}
    \item one self-impedance element $Z_{\text{T},kk}^{\textrm{self}}$ already expressed in (\ref{eq:Z1-chu}),
    \item $(N-1)$ mutual impedance elements  $Z_{\text{T},kn}^{\textrm{mutual}}$ for $n\in\{1,\dots,N\}\backslash\{k\}$ which are given by the coupling model $\mathcal{MC}(d,\beta,\gamma)$ in (\ref{eq:mutual-coupling-SISO}).
\end{itemize}

\noindent To find the asymptotically optimum spacing-to-antenna-size ratio $(\delta/a)^{\tiny{\starletfill}}$ of tightly-coupled massive MIMO, we let $N \rightarrow \infty$. Without loss of generality, we also consider a uniform current excitation of the transmit array, i.e., $\bm{i}_{\text{T}}(f) \propto \mathbf{1}_{N\times 1}$ (i.e., broadside beamformer). By doing so, the $k$th voltage element of $\bm{v}_{\textrm{T}}(f)$ can be written as

\begin{equation}\label{appendix:v=Zi}
    v_{k}(f)= Z^\text{self}_{\textrm{T},kk}(f) + \sum\limits_{\substack{n=-\infty\\n\neq k}}^{+\infty} Z^\text{mutual}_{\textrm{T},kn}(f).
\end{equation}

\noindent We assume that the transmit array is an infinite Chu's CMS array, i.e.,  $Z^\text{self}_{\textrm{T},kk}(f)\triangleq Z^\text{self}_{\textrm{Chu},kk}(f)$ and $Z^\text{mutual}_{\textrm{T},kn}(f) = Z^\text{mutual}_{\textrm{Chu},kn}(f)$.

\noindent The optimum spacing-to-antenna-size ratio $(\delta/a)^{\tiny{\starletfill}}$ maximizes the radiation resistance of the Chu's CMS arrays, and hence cancels out the reactive part of the transmit array impedance. The latter must be matched to the real internal resistance $R$ of the voltage generator $\bm{v}_{\textrm{T}}(f)$.  This is equivalent to cancelling out $\Im\{v_k(f)\}$ for all $k$ in (\ref{appendix:v=Zi}), thereby leading to the following condition for tight coupling:
\begin{equation}\label{appendix:reactive-part=0}
    \Im\left\{Z^\text{self}_{\text{Chu},kk}(f)\right\} + \sum_{\substack{\ell=-\infty\\\ell\neq k}}^{+\infty} \Im\left\{Z^\text{mutual}_{\text{Chu},k\ell}(f)\right\} = 0.
\end{equation}
To further simplify (\ref{appendix:reactive-part=0}), we first find the real and imaginary parts of the $k$th self-impedance $Z^\text{self}_{\text{Chu},kk}(f)$ given in (\ref{eq:Z1-chu}) as follows:
\begin{subequations}
    \begin{align}
        \Re\{Z^\text{self}_{\text{Chu},kk}(f)\} &= \frac{(2\pi fa)^2R}{(2\pi fa)^2 + c^2}~ [\Omega],\label{appendix:real-Chu-self}\\
        \Im\{Z^\text{self}_{\text{Chu},kk}(f)\} &=-\frac{c^3R}{(2\pi fa)((2\pi fa)^2+c^2)}~[\Omega]\label{appendix:imaginary-Chu-self}.
    \end{align}
\end{subequations}
\noindent For colinear arrays, we set $\beta=0$, $\gamma=\pi$ and $d=|\ell\,\delta|$ for $\ell\in\mathbb{Z} \backslash \{0\}$ in the coupling model $\mathcal{MC}(d,\beta,\gamma)$ established in (\ref{eq:mutual-coupling-SISO}). This yields the corresponding mutual impedance $Z^\text{mutual}_{\text{Chu},\textrm{colinear},k\ell}(f)\triangleq Z^\text{mutual}_{\text{Chu},k\ell}(f)\big|_{\beta=\gamma=0}$ whose  imaginary part is given by:

\begin{equation}\label{appendix:imaginary-Chu-mutual}
\begin{aligned}[b]
    &\Im\{Z^\text{mutual}_{\text{Chu},\textrm{colinear},k\ell}(f)\} \\
    &=3\,\sqrt{\Re\big\{Z^{\textrm{self}}_{\text{T},kk}\big\}\,\Re\big\{Z^{\textrm{self}}_{\text{R},kk}\big\}}\,\left[\frac{\cos(k_0|\delta\ell|)}{(k_0|\delta\ell|)^3} + \frac{\sin(k_0|\delta\ell|)}{(k_0|\delta\ell|)^2}\right],\\
        &= \frac{3\,(2\pi fa)^2R}{(2\pi fa)^2 + c^2} \left[\frac{\cos(k_0|\delta\ell|)}{(k_0|\delta\ell|)^3} + \frac{\sin(k_0|\delta\ell|)}{(k_0|\delta\ell|)^2}\right],
        \end{aligned}
\end{equation}
where the last equality follows from (\ref{appendix:real-Chu-self}) because both transmit and receive arrays are Chu's CMS arrays, i.e., $Z^{\textrm{self}}_{\text{T},kk} = Z^{\textrm{self}}_{\text{R},kk} = Z^{\textrm{self}}_{\text{Chu},kk}$. Now, by injecting (\ref{appendix:imaginary-Chu-self}) and (\ref{appendix:imaginary-Chu-mutual}) back into (\ref{appendix:reactive-part=0}), the tight coupling condition for colinear arrays becomes:
\begin{equation}\label{appendix:equality-colinear-tight-condition}
\begin{aligned}
    &\frac{c^3R}{(2\pi fa)((2\pi fa)^2+c^2)} \\
    &= \frac{3\,(2\pi fa)^2R}{(2\pi fa)^2 + c^2} \left(\sum_{\substack{\ell=-\infty\\\ell\neq0}}^{+\infty} \frac{\cos(k_0|\delta\ell|)}{(k_0|\delta\ell|)^3} + \sum_{\substack{\ell=-\infty\\\ell \neq 0}}^{+\infty} \frac{\sin(k_0|\delta\ell|)}{(k_0|\delta\ell|)^2}\right),\\
    &= \frac{3\,(2\pi fa)^2R}{(2\pi fa)^2 + c^2} \left(2\sum_{\substack{\ell=1}}^{+\infty} \frac{\cos(k_0\delta\ell)}{(k_0|\delta\ell|)^3} +2 \sum_{\substack{\ell=1}}^{+\infty} \frac{\sin(k_0|\delta\ell|)}{(k_0\delta\ell)^2}\right),\\
    &= \frac{3\,(2\pi fa)^2R}{(2\pi fa)^2 + c^2} \Bigg(\frac{\text{Li}_3\left(e^{-\text{j}k_0\delta}\right) + \text{Li}_3\left(e^{\text{j}k_0\delta}\right)}{(k_0\delta)^3} \\
    &\hspace{3cm}+ \frac{\text{j}\left(\text{Li}_2\left(e^{-\text{j}k_0\delta}\right) - \text{Li}_2\left(e^{\text{j}k_0\delta}\right)\right)}{(k_0\delta)^2}\Bigg),
\end{aligned}
\end{equation}
where $\text{Li}_{s}(\cdot)$ denotes the polylogarithm function of order $s$. After basic algebraic manipulations, we show that (\ref{appendix:equality-colinear-tight-condition}) reduces to:
\begin{equation}\label{appendix:equality-colinear-tight-condition-2}
\begin{aligned}[b]
    \frac{1}{3}\left(\frac{\delta}{a}\right)^3 &= \text{Li}_3(e^{-\text{j}k_0\delta}) + \text{Li}_3(e^{\text{j}k_0\delta}) \\
    & \hspace{1cm} + \text{j}k_0\delta\left(\text{Li}_2\left(e^{-\text{j}k_0\delta}\right) - \text{Li}_2\left(e^{\text{j}k_0\delta}\right)\right).
\end{aligned}
\end{equation}
Then, taking the limit in both sides of (\ref{appendix:equality-colinear-tight-condition}) as $k_0\delta \rightarrow 0$ (i.e., in the limit of quasi-continuous aperture) yields:
\begin{equation}\label{appendix:equality-colinear-tight-condition-3}
    \frac{1}{3}\left(\frac{\delta}{a}\right)^3 = 2 \,\zeta(3),
\end{equation}
where $\zeta(\cdot)$ is the Riemann zeta function. Finally, solving (\ref{appendix:equality-colinear-tight-condition-3}) for the spacing-to-antenna-size ratio $\delta/a$ leads to the desired expression (\ref{eq:result-1-formula}) in Result \ref{result:condition-tight-coupling}.
\end{appendices}

\bibliographystyle{IEEEtran}
\bibliography{IEEEabrv,references}

\end{document}